\documentclass[
superscriptaddress,
twocolumn,
showpacs,preprintnumbers,
amsmath,amssymb,
aps,
prb,
]{revtex4-2}
\usepackage[colorlinks,
linkcolor=blue, urlcolor=blue, anchorcolor=blue, citecolor=blue]{hyperref}
\usepackage{graphicx}
\usepackage{dcolumn}
\usepackage{bm}
\usepackage{float}
\usepackage{multirow}
\usepackage{longtable}
\usepackage{amsmath}
\usepackage{amssymb}
\usepackage{xcolor}
\usepackage{lipsum}  
\usepackage[normalem]{ulem}
\usepackage{titlesec}

\newcommand{\angstrom}{\textup{\AA}}
\begin{document}


\title{Understanding the Ising zigzag antiferromagnetism of FePS$_3$ and FePSe$_3$ monolayers} 


\author{Ke Yang}
\affiliation{College of Science, University of Shanghai for Science and Technology, Shanghai 200093, China}
\affiliation{Laboratory for Computational Physical Sciences (MOE),
 State Key Laboratory of Surface Physics, and Department of Physics,
 Fudan University, Shanghai 200433, China}

\author{Yueyue Ning}
\affiliation{College of Science, University of Shanghai for Science and Technology, Shanghai 200093, China}

\author{Yuxuan Zhou}
 \affiliation{Laboratory for Computational Physical Sciences (MOE),
 State Key Laboratory of Surface Physics, and Department of Physics,
  Fudan University, Shanghai 200433, China}
\affiliation{Shanghai Qi Zhi Institute, Shanghai 200232, China}

\author{Di Lu}
\affiliation{Laboratory for Computational Physical Sciences (MOE),
State Key Laboratory of Surface Physics, and Department of Physics,
 Fudan University, Shanghai 200433, China}
\affiliation{Shanghai Qi Zhi Institute, Shanghai 200232, China}

\author{Yaozhenghang Ma}
 \affiliation{Laboratory for Computational Physical Sciences (MOE),
 State Key Laboratory of Surface Physics, and Department of Physics,
  Fudan University, Shanghai 200433, China}
\affiliation{Shanghai Qi Zhi Institute, Shanghai 200232, China}

\author{Lu Liu}
 \affiliation{Laboratory for Computational Physical Sciences (MOE),
 State Key Laboratory of Surface Physics, and Department of Physics,
  Fudan University, Shanghai 200433, China}
\affiliation{Shanghai Qi Zhi Institute, Shanghai 200232, China}

\author{Shengli Pu}
\affiliation{College of Science, University of Shanghai for Science and Technology, Shanghai 200093, China}

 \author{Hua Wu}
 \email{Corresponding author. wuh@fudan.edu.cn}
 \affiliation{Laboratory for Computational Physical Sciences (MOE),
  State Key Laboratory of Surface Physics, and Department of Physics,
  Fudan University, Shanghai 200433, China}
 \affiliation{Shanghai Qi Zhi Institute, Shanghai 200232, China}
 \affiliation{Shanghai Branch, Hefei National Laboratory, Shanghai 201315, China}

\date{\today}

\begin{abstract}

Transition metal phosphorous trichalcogenides represent a class of van der Waals magnetic materials ideal for exploring two-dimensional (2D) magnetism. This study investigates the spin-orbital states of FePS$_3$ and FePSe$_3$ monolayers and the origin of their Ising zigzag antiferromagnetism (AFM), using density functional calculations, crystal field level diagrams, superexchange analyses, and parallel tempering Monte Carlo (PTMC) simulations. Our calculations show that under the trigonal elongation of the FeS$_6$ (FeSe$_6$) octahedra, the $e_g^\pi$ doublet of the Fe $3d$ crystal field levels lies lower than the $a_{1g}$ singlet by about 108 meV (123 meV), which is much larger than the strength of Fe $3d$ spin-orbit coupling (SOC). Then, the half-filled minority-spin $e_g^\pi$ doublet of the high-spin Fe$^{2+}$ ions ($d^{5\uparrow,1\downarrow}$) splits by the SOC into the lower $L_{z+}$ and higher $L_{z-}$ states. The spin-orbital ground state $d^{5\uparrow}$$L_{z+}^{1\downarrow}$ formally with $S_z$ = 2 and $L_z$ = 1 gives the large $z$-axis spin/orbital moments of 3.51/0.76 $\mu_{\rm B}$ (3.41/0.67 $\mu_{\rm B}$) for FePS$_3$ (FePSe$_3$) monolayer, and both the moments are reduced by the strong (stronger) Fe $3d$ hybridizations with S $3p$ (Se $4p$) states. As a result, FePS$_3$ (FePSe$_3$) monolayer has a huge perpendicular single-ion anisotropy (SIA) energy of 19.4 meV (14.9 meV), giving an Ising-type magnetism. Moreover, via the maximally localized Wannier functions, we find that the first nearest neighboring (1NN) Fe-Fe pair has large hopping parameters in between some specific orbitals, and so does the 3NN Fe-Fe pair. In contrast, the 2NN Fe-Fe pair has much smaller hopping parameters and the 4NN Fe-Fe pair has negligibly small ones. Then, a combination of those hopping parameters and the superexchange picture can readily explain the computed strong 1NN ferromagnetic coupling and the strong 3NN antiferromagnetic one but the relatively much smaller 2NN antiferromagnetic coupling. Furthermore, our PTMC simulations give $T_{\rm N}$ of 119 K for FePS$_3$ monolayer and well reproduce its experimental Ising zigzag AFM, and also predict for FePSe$_3$ monolayer the same magnetic structure with a close or even higher $T_{\rm N}$.
\end{abstract}

\maketitle


\section*{I. Introduction}

The study of two-dimensional (2D) magnetic materials has seen a surge in interest~\cite{Novoselov2004,Geim2007,Li2019,Song2019,Li2020,Burch2018,Song2018} following the discovery of ferromagnetic (FM) behavior in the CrI$_3$ monolayer~\cite{Huang2017} and the Cr$_2$Ge$_2$Te$_6$ bilayer~\cite{Gong2017} in 2017. In line with the Mermin-Wagner theorem~\cite{Mermin1966}, the key to establishing magnetic order in 2D materials is magnetic anisotropy (MA). 
For instance, both the CrI$_3$ monolayer~\cite{Huang2017} and Cr$_2$Ge$_2$Te$_6$ bilayer~\cite{Gong2017} exhibit weak out-of-plane anisotropy due to the spin-orbit coupling (SOC) of heavy ligand $p$ orbitals and their hybridization with closed Cr$^{3+}$ $t_{2g}^3$ shell~\cite{Wang_2016,Lado2017,Xu2018,Kim2019}. In stark contrast, the VI$_3$ monolayer exhibits giant single-ion anisotropy (SIA) of 16 meV per V atom associated with the open V$^{3+}$ $t_{2g}^{2}$ shell and its SOC effects~\cite{Yang2020,Zhao2021,DeVita2022}, and the experimental large orbital moment is 0.6 $\mu_{\rm B}$~\cite{Lin2021,Hovancik2023}. As the MA is essential to the 2D magnetism, it is desirable to have a large orbital moment, a giant SIA and thus Ising-type magnetism~\cite{Gibertini2019,Mak2019,Gong2019}.

\begin{figure}[t]
	\centering 
	\includegraphics[width=9cm]{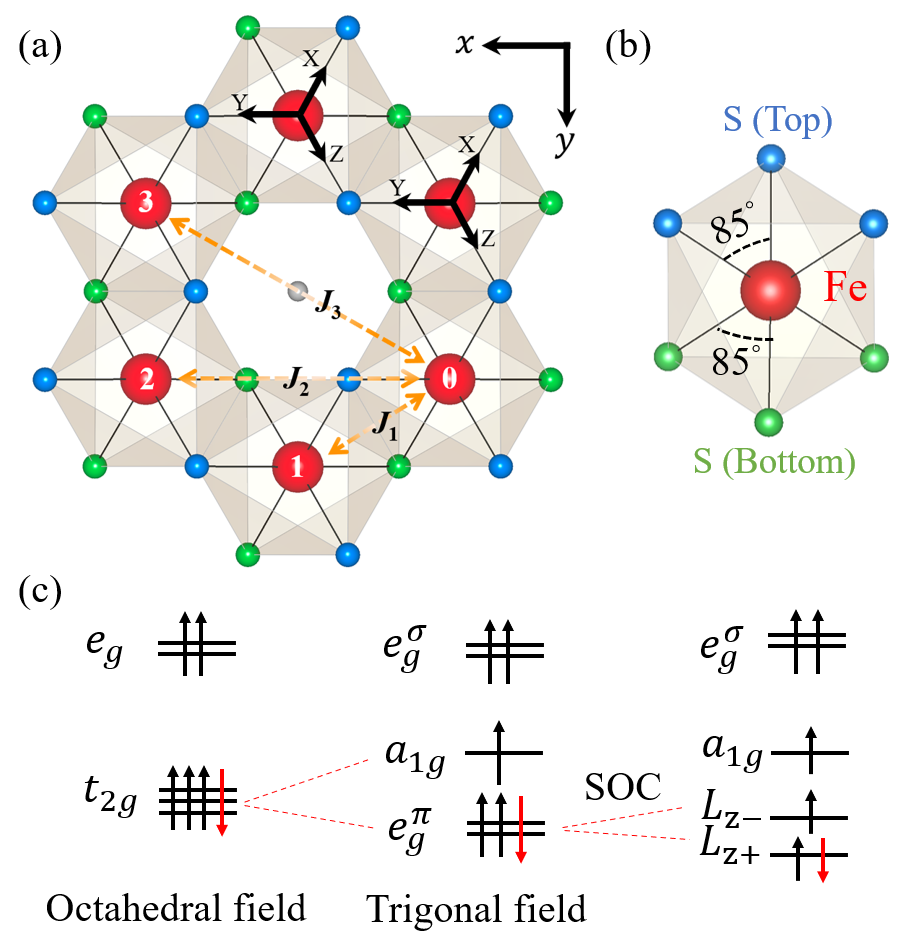}
	\centering
	\caption{(a) The atomic structure of FeP$X_3$ ($X$ = S, Se) monolayer, with the P-P dimer located vertically across the center of the honeycomb formed by Fe ions, and (b) the edge-shared Fe$X_6$ ($X$ = S, Se) octahedron. (c) The Fe$^{2+}$ 3$d^6$ spin-orbital states with $S$ = 2 in the local octahedral but global trigonal crystal field.
}
	\label{structure}
\end{figure}

Transition metal phosphorous trichalcogenides are a class of van der Waals layered materials, and among them, FePS$_3$ and FePSe$_3$ bulk materials are zigzag antiferromagnetic (AF) semiconductors~\cite{Wiedenmann1981,Lee2016,Coak2019,Zheng2019} with close N$\acute{\rm e}$el temperatures ($T_{\rm N}$) of 123 K~\cite{Joy1992} and 119 K~\cite{Wiedenmann1981}, respectively. Both compounds contain the Fe$^{2+}$ ion with $S$ = 2 and exhibit their respective effective magnetic moment of 5.23 $\mu_{\rm B}$ for FePS$_3$ ~\cite{Joy1992} and 4.90 $\mu_{\rm B}$ for FePSe$_3$~\cite{Wiedenmann1981}. 
Note that taking into account a covalent reduction, a spin-only $S$ = 2 state would have the effective magnetic moment less than 4.9 $\mu_{\rm B}$ ($\sqrt{g^{2}_{s}S(S+1)}$ = $\sqrt{2^2\cdot2\cdot3}$ $\approx$ 4.9). Therefore, the above effective moments larger than or equal to 4.9 $\mu_{\rm B}$ imply a contribution of additional orbital moments~\cite{Joy1992}.

Recently, FePS$_3$ has been successfully exfoliated to a monolayer, which maintains the zigzag AF ordering with $T_{\rm N}$ = 118 K~\cite{Lee2016,zhang2021,Wang2016}. This $T_{\rm N}$ remains almost independent of thickness, from bulk to the monolayer limit, indicating the predominance of a strong Ising-type magnetism~\cite{Lee2016}. 
Several theoretical works ~\cite{Chittari2016,Olsen2021,Zhang_2021,Kim2021,Amirabbasi2023,Li2024} have confirmed the zigzag AF ground state, showing a strong first nearest neighboring (1NN) FM coupling and a strong 3NN AF one but a much smaller 2NN AF coupling. 
So far, the Ising magnetism has been confirmed~\cite{Lee2016,Joy1992,Kim2021}, however, the corresponding electronic structure and spin-orbital state remain less clear.
Therefore, here we provide an insight into the Ising magnetism of FePS$_3$ and FePSe$_3$ monolayers. Moreover, we use the $i$NN ($i$ = 1-4) hopping parameters, derived from the Wannier functions, to explain the competitive 1NN FM and 3NN AF couplings in determining the zigzag AFM. Then, our PTMC simulations yield a rational $T_{\rm N}$ value.

In FePS$_3$ and FePSe$_3$ monolayers, Fe$^{2+}$ ions in the local octahedral coordination are influenced by the global trigonal crystal field, see Fig. \ref{structure}. This field splits the $t_{2g}$ triplet into a higher $a_{1g}$ singlet and a lower $e_g^{\pi}$ doublet due to the elongation of the octahedra along the global $z$-axis, as illustrated in Figs. \ref{structure}(b) and \ref{structure}(c). The crystal field splitting between the $a_{1g}$ singlet and $e_g^{\pi}$ doublet is calculated to be 108 (123) meV for FePS$_3$ (FePSe$_3$) as seen below. After a direct comparison of different spin-orbital states in the following calculations, we find the $3d^{5\uparrow}L_{z+}^{1\downarrow}$ ground state with the formal $S_z = 2$ and $L_z = 1$, as seen in Fig. \ref{structure}(c). Then, the consequent big perpendicular orbital moment and huge SIA readily account for the experimental Ising magnetism. Moreover, our present work consistently explains the Ising type zigzag AFM of FePS$_3$ and FePSe$_3$ monolayers, using the first-principles calculations of their spin-orbital states and the superexchange parameters, the hopping parameters derived from Wannier functions, and the PTMC simulations. Furthermore, we predict that their $T_{\rm N}$ would be enhanced under a compressive strain.

\section*{II. Computational Details}

Density functional theory (DFT) calculations are performed using the Vienna $Ab$ $initio$ Simulation Package (VASP)~\cite{Kresse1993}. The generalized gradient approximation (GGA) proposed by Perdew, Burke, and Ernzerhof (PBE)~\cite{Perdew1996} is used to describe the exchange-correlation potential. The optimized lattice constants $a$ = $b$ = 5.93 (6.28) $\angstrom$ for FePS$_3$ (FePSe$_3$) monolayer are close to the experimental bulk values of 5.95~\cite{Ouvrard1985} (6.27~\cite{Wiedenmann1981}) $\angstrom$. 
A 20 $\angstrom$ thick slab is used to model FeP$X_3$ ($X$ = S, Se) monolayer. The kinetic energy cutoff is set to 450 eV.
The total energies and atomic forces converge to 10$^{-5}$ eV and 0.01 eV/$\angstrom$. A Monkhorst-Pack $k$-mesh of 6$\times$6$\times$1 (6$\times$3$\times$1) is used for 1$\times$1 unit cell (1$\times$$\sqrt{3}$ supercell). The setup of kinetic energy and $k$-point sampling is carefully tested, as seen in Table S1 in Supplemental Material (SM)~\cite{SM}.

To describe the correlation effect of the localized Fe 3$d$ electrons, we employ the GGA + $U$ approach~\cite{Anisimov1997}. The Hubbard $U$ values are calculated, through the constrained random phase approximation~\cite{CRPA}, to be 3.6 eV for FePS$_3$ monolayer and 2.7 eV for FePSe$_3$, with the common value of Hund's exchange $J_{\rm H}$ = 0.9 eV. 
Both the $U$ and $J_{\rm H}$ are included in the following GGA + $U$ + $J_{\rm H}$ (normally referred to as GGA + $U$) calculations. The reduction of the $U$ values from FePS$_3$ to FePSe$_3$ could mainly be due to an enhanced Coulomb screening effect resulting from a stronger Fe $3d$-Se $4p$ hybridization. 
Moreover, the stronger Fe $3d$-Se $4p$ hybridizations will affect the hopping integrals and superexchange, as seen in subsection III E.
Note that $J_{\rm H}$ is actually the difference of the energies of electrons with different spins or orbitals on the same atomic shell, and therefore, $J_{\rm H}$ is almost not screened and not modified when going from an atom to a solid. It is almost a constant for a given element and is typically 0.8-1.0 eV for a 3$d$ transition metal~\cite{Khomskii_2014}.
We also test the common $U$ = 4 eV for both FePS$_3$ and FePSe$_3$ monolayers (see Table S2 and Fig. S1 in SM~\cite{SM}) and find that some quantitative change of the results does not affect our conclusions. To figure out the ground state among a set of spin-orbital states, the occupation number matrices are controlled in our calculations using the open-source software developed by Watson~\cite{Allen2014}. The SOC is included in our calculations using the second-variational method with scalar relativistic wave functions.

The hopping parameters are obtained from maximally localized Wannier functions (MLWFs) using the Wannier90~\cite{Mostofi2008,Nicola2012}.
Moreover, we perform PTMC~\cite{PTMC} simulations to estimate the $T_{\rm N}$ of FePS$_3$ and FePSe$_3$ monolayers on a 12$\times$12$\times$1 spin matrix with periodic boundary conditions, and the number of replicas is set to 112. 
A similar result is obtained with larger supercells. During the simulation step, each spin is rotated randomly in the three-dimensional space. The spin dynamical process is studied by the classical Metropolis methods~\cite{Metropolis1949}.

We adopt the global coordinate system with the $z$-axis along the local [111] direction of the Fe$X_6$ ($X$ = S, Se) octahedra, see Fig. ~\ref{structure}(a) for the local $XYZ$ axes and global $xyz$ axes. The eigenwave function in the local octahedral but global trigonal crystal field, under the global $xyz$ coordinate system, can be expressed as
%
%

\begin{equation} \label{eq: 1}
    \begin{gathered}
        \left|a_{1 g}\right\rangle=\left|3 z^{2}-r^2\right\rangle\\
        \left|e_{g_1}^{\pi}\right\rangle=\sqrt{\frac{2}{3}}\left|x^{2}-y^{2}\right\rangle-\sqrt{\frac{1}{3}}\left|xz\right\rangle \\
        \left|e_{g_2}^{\pi}\right\rangle=\sqrt{\frac{1}{3}}\left| yz\right\rangle+\sqrt{\frac{2}{3}}\left|xy\right\rangle \\
        \left|e_{g_1}^\sigma\right\rangle=\sqrt{\frac{1}{3}}\left|x^{2}-y^{2}\right\rangle+\sqrt{\frac{2}{3}}\left|xz\right\rangle \\
        \left|e_{g_2}^\sigma\right\rangle=\sqrt{\frac{2}{3}}\left|yz\right\rangle-\sqrt{\frac{1}{3}}\left|xy\right\rangle
    \end{gathered}
\end{equation}
Considering the crystal field splitting and SOC effect, the half-filled down-spin $e_{g}^{\pi}$ doublet of the Fe$^{2+}$ 3$d^6$ high-spin state would carry an unquenched orbital moment characterized by the effective orbital momentum $L = 1$, see Fig. ~\ref{structure}(c). Their eigenwave functions are expressed as
\begin{equation} \label{eq: 2}
	\begin{aligned}
		&\left|L_{z\pm}\right\rangle=\frac{1}{\sqrt{2}}\left(\left|e_{g_1}^\pi\right\rangle\pm i\left|e_{g_2}^\pi\right\rangle\right) 
	\end{aligned}
\end{equation}
where $L_{z}$ represents the projection of orbital moment along the $z$-axis, and $\pm$ stands for the $L_z$ = $\pm$1.

\section*{III. Results and Discussion}

\subsection*{A. The Fe$^{2+}$ high-spin $S$ = 2 state}

\begin{figure}[t]
	\includegraphics[width=7cm]{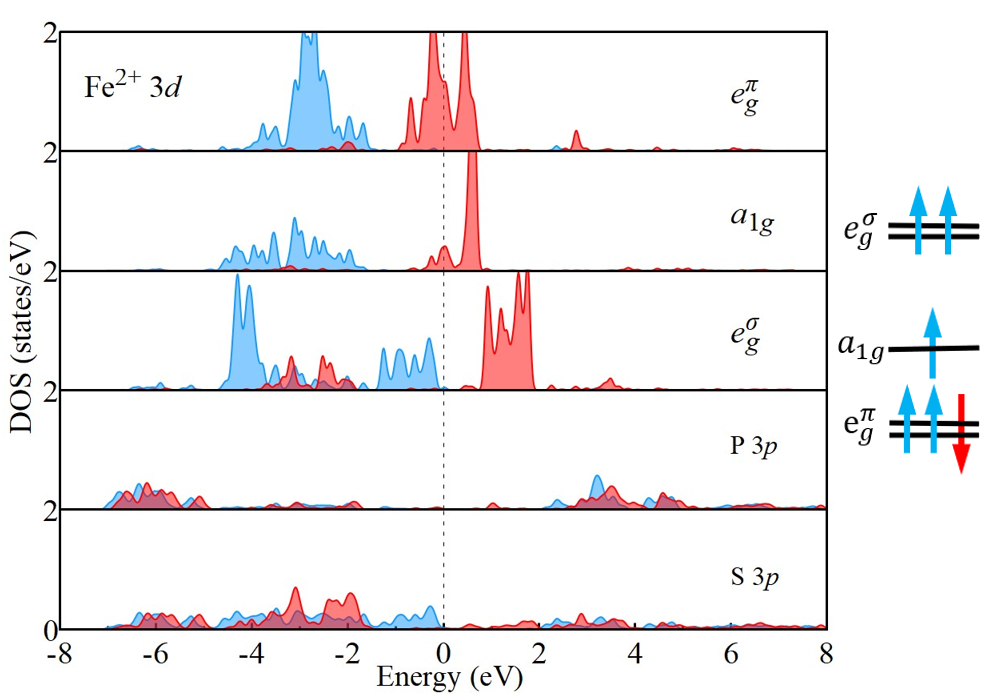}
	\centering
	\caption{Density of states (DOS) of FePS$_3$ monolayer by GGA calculation. The Fermi level is set at zero energy. The blue (red) curves stand for the up (down) spin channel. 
	}
	\label{GGAS}
\end{figure}

We initially focus on FePS$_3$ monolayer, for which experimental results are available for comparison~\cite{Lee2016,zhang2021,Wang2016}.
To see the crystal field effect, exchange splitting, electron correlation, and the crucial SOC effects, we present and discuss below the spin-polarized GGA and GGA + SOC + $U$ calculations.
Firstly, we perform spin-polarized GGA calculations for the FM state to investigate the crystal field effect and the charge-spin state of the FePS$_3$ monolayer. 
As shown in Fig.~\ref{GGAS}, the octahedral $t_{2g}$-$e_g$ crystal field splitting is about 1 eV in good agreement with the experiment~\cite{Mertens_2023}, and in the global trigonal crystal field, the $t_{2g}$ triplet further splits into the $a_{1g}$ singlet and $e_g^{\pi}$ doublet.
The five up-spin 3$d$ orbitals are fully occupied, while the down-spin $e_g^{\pi}$ doublet is half-filled, leading to the Fe$^{2+}$ 3$d^{5\uparrow}(e_g^{\pi})^{1\downarrow}$ configuration. 
In comparison with the Fe 3$d$ orbitals around the Fermi level, however, the P 3$p$ orbitals have little contribution around the Fermi level. Instead, there is a large bonding-antibonding split (about --6 eV vs 3 eV both relative to the Fermi level), which arises from the P-P dimerization. In contrast, the S 3$p$ states have strong hybridization with Fe 3$d$ orbitals, and they have a large and broad contribution in the energy range from --6 eV to the Fermi level. Therefore, it is undoubted that the S 3$p$ states would play a vital role in the superexchange interactions mediating the magnetic couplings in the Fe sublattice.

Moreover, FePS$_3$ monolayer exhibits a total spin moment of 3.97 $\mu_{\rm B}$/f.u. for the FM state, suggesting the formal Fe$^{2+}$ $S$ = 2 high-spin state. The Fe$^{2+}$ ion displays a local spin moment of 3.37 $\mu_{\rm B}$. Owing to the Fe $3d$-S $3p$ hybridization, each sulfur ion gets spin-polarized and has a local spin moment of 0.11 $\mu_{\rm B}$, and an additional spin moment of 0.28 $\mu_{\rm B}$/f.u. appears in the interstitial region. 
To confirm the high-spin $S=2$ ground state of the Fe$^{2+}$ ion, we also compute the low-spin $S=0$ state and find it to be 796 meV/f.u. unstable against the high-spin ground state. Note that it is the magnitude of the $t_{2g}$-$e_g$ octahedral crystal field splitting ($\Delta_{\rm cf}$) relative to $J_{\rm H}$ which determines the spin state of the Fe$^{2+}$ ion. To make a crude estimate: the $S$ = 2 state (3$d^{5\uparrow}t_{2g}^{1\downarrow}$) carries the Hund's coupling energy of $-$10$J_{\rm H}$ plus 2$\Delta_{\rm cf}$ (the crystal field excitation energy of two electrons on the $e_g$ orbitals), whereas the $S$ = 0 state ($t_{2g}^{3\uparrow 3\downarrow}$) has a total stabilization energy of $-$6$J_{\rm H}$. 
Therefore, a critical value for a high-spin to low-spin transition is $\Delta_{\rm cf}>$ 2$J_{\rm H}$ = 1.8 eV.
As seen in Fig. \ref{GGAS}, the $t_{2g}$-$e_g$ crystal field splitting $\Delta_{\rm cf}$ is about 1 eV and much smaller than the critical value of 1.8 eV. Therefore, FePS$_3$ monolayer is well stabilized in the high-spin $S$ = 2 ground state.

As seen above, FePS$_3$ monolayer exhibits the Fe$^{2+}$ 3$d^{5\uparrow}(e_g^{\pi})^{1\downarrow}$ high-spin state. 
The $e_g^{\pi}$ doublet has a lower crystal field energy than the $a_{1g}$ singlet, which accords with the elongation of the FeS$_6$ octahedra along the $z$-axis as depicted in Fig.~\ref{structure}(b), where the marked S-Fe-S bond angles of 85 degrees deviate from the ideal ones of 90 degrees. As a result, the $a_{1g}$ singlet rises up in the crystal field energy. As seen below, the half filling of the down-spin $e_g^{\pi}$ doublet is crucial for the Ising magnetism of FePS$_3$. When the SOC is included ($\zeta \overrightarrow{l} \cdot \overrightarrow{s}$), the $e_g^{\pi}$ doublet would split into the $L_{z+} = +1$ and $L_{z-} = -1$ states as expressed in Eq. ~\ref{eq: 2}. 

\begin{figure}[t]
	\includegraphics[width=8cm]{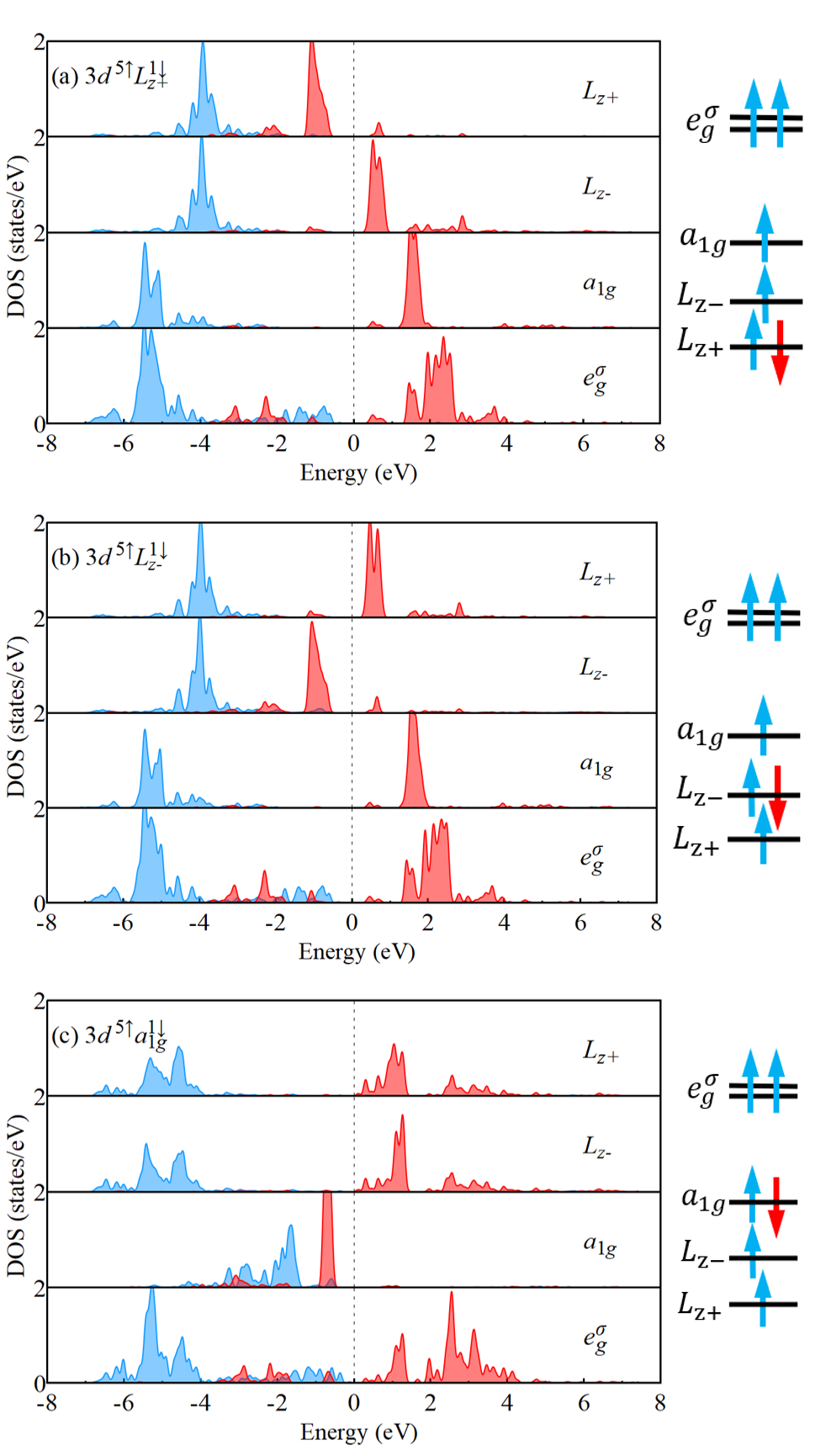}
	\centering
	\caption{The DOS results of FePS$_3$ monolayer in (a) the 3$d^{5\uparrow}$$L_{z+}^{1\downarrow}$ ground state, (b) 3$d^{5\uparrow}$$L_{z-}^{1\downarrow}$ and (c) 3$d^{5\uparrow}$$a_{1g}^{1\downarrow}$ states by the GGA + SOC + $U$ calculations, and the corresponding crystal field level diagrams. The blue (red) curves stand for the up (down) spin channel. The Fermi level is set at zero energy.}
	\label{GGA_U_SOC_DOS}
\end{figure}

\subsection*{B. The $L_z = 1$ ground state and Ising magnetism}

Then, to investigate the effects of SOC and electron correlation, we perform GGA + SOC + $U$ calculations.
The obtained insulating solution has a total spin moment of 4 $\mu_{\rm B}$/f.u. for the FM state and a local spin moment of 3.55 $\mu_{\rm B}$ for the Fe ion, reinforcing the formal Fe$^{2+}$ $S = 2$ state. Additionally, we observe a large orbital moment of 0.73 $\mu_{\rm B}$ on the Fe$^{2+}$ ion, in line with the splitting of the $e_g^{\pi}$ doublet due to SOC, resulting in the 3$d^{5\uparrow}L_{z+}^{1\downarrow}$ configuration.
As depicted in Fig. \ref{GGA_U_SOC_DOS}(a), the five up-spin 3$d$ orbitals are fully occupied. The single down-spin electron occupies the lower-energy $L_{z+}$ orbital, leaving the higher-energy $L_{z-}$ orbital empty, resulting in a semiconductor with a band gap of 0.7 eV. This leads to the 3$d^{5\uparrow}L_{z+}^{1\downarrow}$ state with a large orbital moment along the $z$-axis.

\begin{figure}[t]
	\includegraphics[width=8cm]{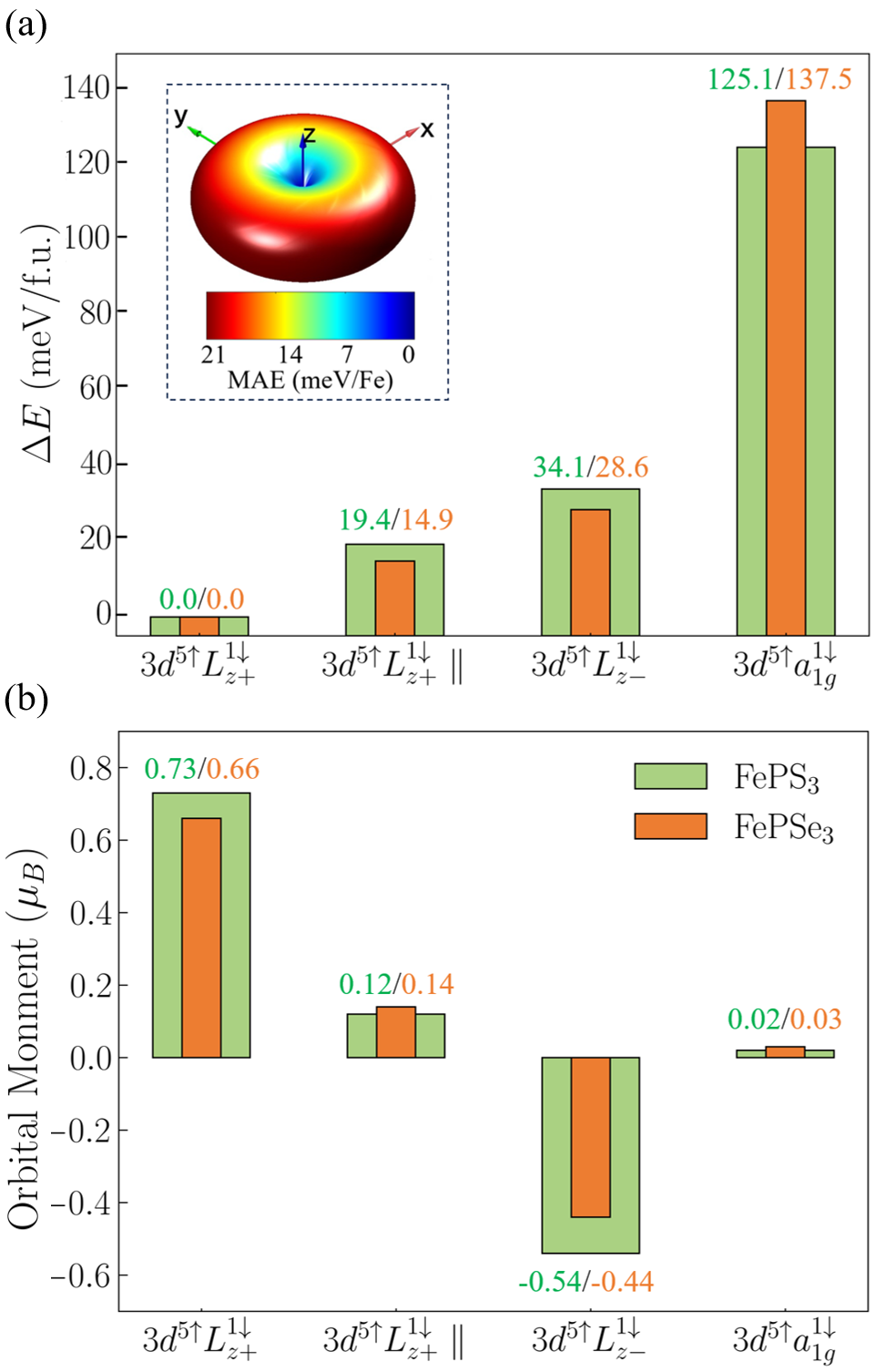}
	\centering
	\caption{(a) The relative total energies $\Delta$$\it{E}$ (meV/f.u.), and (b) orbital moments for FePS$_3$ and FePSe$_3$ monolayers in different spin-orbital states by GGA + SOC + $U$ calculations. The symbol $\Vert$ in the state labeling marks the in-plane magnetization, in comparison with other states with out-of-plane magnetization.   
}
	\label{states}
\end{figure}

The band gap of 0.7 eV between the occupied $L_{z+}$ and unoccupied $L_{z-}$ states is mainly due to the electron correlation effect, and it significantly exceeds the typical SOC effect. To evaluate the spin-orbital excitation energy such as the $L_{z+}$/$L_{z-}$ orbital splitting by SOC, we focus on the computed total energy differences between the 3$d^{5\uparrow}L_{z+}^{1\downarrow}$ and 3$d^{5\uparrow}L_{z-}^{1\downarrow}$ configurations, other than relying on the DOS results. For this assessment, we perform GGA + SOC + $U$ calculations, enabling a direct comparison between these configurations, initialized via the occupation density matrix over the eigenorbitals.

As depicted in Fig. \ref{GGA_U_SOC_DOS}(b), the Fe$^{2+}$ five up-spin 3$d$ orbitals are fully occupied, and the single down-spin electron occupies the $L_{z-}$ orbital, forming the 3$d^{5\uparrow}L_{z-}^{1\downarrow}$ spin-orbital state. The Fe$^{2+}$ ion now has a local spin moment of 3.54 $\mu_{\rm B}$, and notably, it exhibits an opposite orbital moment of $-$0.54 $\mu_{\rm B}$, see the results in Fig. \ref{states}. In the 3$d^{5\uparrow}L_{z-}^{1\downarrow}$ state, the up-spin subshell is closed, and the single down-spin electron unfavorably carries a parallel negative orbital moment. As a result, the 3$d^{5\uparrow}L_{z-}^{1\downarrow}$ state rises in the SOC energy against the 3$d^{5\uparrow}L_{z+}^{1\downarrow}$ state by $\Delta E_{\rm SOC}$ = $\zeta$($\Delta l_z$)$s_z$ = $\zeta\cdot (0.73+0.54)\cdot 1/2$, which is 34.1 meV/f.u. as seen in Fig. \ref{states}. Then, here the SOC parameter $\zeta$ is estimated to be 53.7 meV. As the $\zeta$ parameter for the Fe$^{2+}$ ion is typically around 50-60 meV, the present agreement reflects the good accuracy of our calculations.

When studying the SOC effect, it is important to compare the SOC with the crystal field splitting. As the $e_g$-$t_{2g}$ crystal field splitting of about 1 eV is one order of magnitude stronger than the SOC, the $e_g$ doublet is irrelevant when dealing with the SOC. The $a_{1g}$-$e_{g}^{\pi}$ splitting within the $t_{2g}$ triplet is thus of concern. For this purpose, we also stabilize the 3$d^{5\uparrow}a_{1g}^{1\downarrow}$ state in our calculations, see Fig. \ref{GGA_U_SOC_DOS}(c), and then compare it with the above 3$d^{5\uparrow}L_{z+}^{1\downarrow}$ state. Owing to the singlet nature of the $a_{1g}$ orbital, the 3$d^{5\uparrow}a_{1g}^{1\downarrow}$ state has only a tiny orbital moment of 0.02 $\mu_{\rm B}$, in addition to the Fe$^{2+}$ spin moment of 3.51 $\mu_{\rm B}$. Our results show that the 3$d^{5\uparrow}a_{1g}^{1\downarrow}$ state lies above the 3$d^{5\uparrow}L_{z+}^{1\downarrow}$ state by $\Delta E$ = 125.1 meV/f.u. (see Fig. \ref{states}), and this value is close to the experimental one of about 120 meV~\cite{Mertens_2023}.
Then the $a_{1g}$-$e_{g}^{\pi}$ trigonal crystal field splitting can be estimated to be $\Delta E$ $-$ $\frac{1}{2}\Delta E_{\rm SOC}$ = 125.1 $-$ $\frac{1}{2}$ $\times$ 34.1 = 108 meV. This $a_{1g}$-$e_{g}^{\pi}$ splitting is nearly three (two) times as large as the $\Delta E_{\rm SOC}$ of 34.1 meV (the $\zeta$ parameter 53.7 meV), and therefore, the $a_{1g}$-$e_{g}^{\pi}$ mixing by the SOC is insignificant, and then we could restrict our discussion of the SOC effect within the half-filled down-spin $e_{g}^{\pi}$ doublet as seen above.

Through the above calculations of the different spin-orbital states, we find that FePS$_3$ monolayer lies in the 3$d^{5\uparrow}L_{z+}^{1\downarrow}$ ground state and carries the high-spin moment of 3.55 $\mu_{\rm B}$ and a big orbital moment of 0.73 $\mu_{\rm B}$ along the $z$-axis. Owing to the SOC coupling, the spin moment is also fixed along the $z$-axis. If the spin moment was rotated into the $xy$ plane, only a small in-plane orbital moment of 0.12 $\mu_{\rm B}$ would appear, and then the SOC energy would be largely lost and this state has a higher total energy than the 3$d^{5\uparrow}L_{z+}^{1\downarrow}$ ground state solution by 19.4 meV/Fe, see Fig. \ref{states}. As a result, this significant SIA defines the huge perpendicular magnetic anisotropy energy (MAE) of 19.4 meV/Fe, which is two or three orders of magnitude stronger than the MAE in CrI$_3$ monolayer~\cite{Xu2018,Zhao2021,Lado2017,Kim2019}. Therefore, FePS$_3$ displays the robust Ising magnetism as experimentally observed~\cite{Lee2016,zhang2021,Wang2016}.

\subsection*{C. The origin of the zigzag AFM}

\begin{figure}[t]
	\includegraphics[width=8cm]{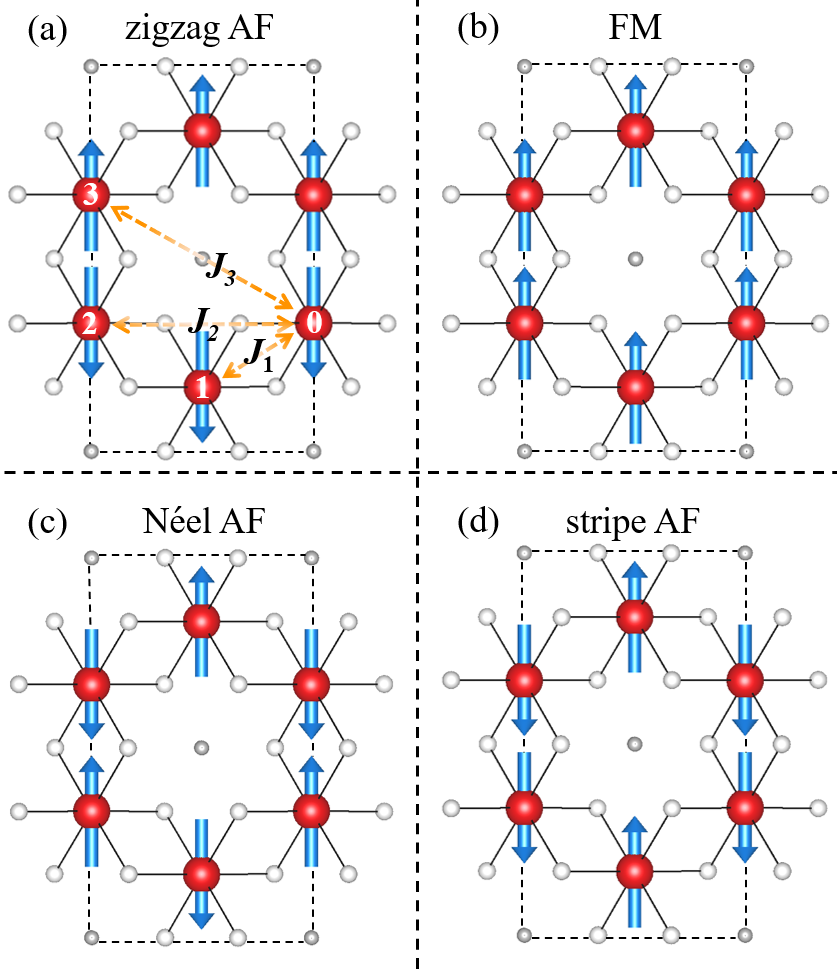}
	\centering
	\caption{The four magnetic structures of FePS$_3$ monolayer marked with three exchange parameters.}
	\label{4Jstructure}
\end{figure}

FePS$_3$ monolayer possesses the $3d^{5\uparrow}L_{z+}^{1\downarrow}$ spin-orbital ground state with $S_z = 2$ and $L_z = 1$. Based on this strong Ising magnetic spin-orbital state, we investigate the experimentally observed zigzag AFM~\cite{Lee2016,zhang2021,Wang2016}. We conduct calculations for three AF states—zigzag AF, N$\acute{\rm e}$el AF, and stripe AF—in addition to the FM state. These states are illustrated in Fig. \ref{4Jstructure}, using a 1$\times\sqrt{3}$ supercell. Our results indicate that the zigzag AF state is most favorable (with a band gap of 1.2 eV, not shown) and exhibits the lowest total energy compared to the other three magnetic states by 17.1-36.7 meV/f.u., as shown in Table \ref{tb3J}. This result confirms the experimentally observed zigzag AFM~\cite{Lee2016,zhang2021,Wang2016}. To seek the origin of the zigzag AF ground state, we identify three exchange parameters: 1NN Fe$_0$-Fe$_1$ ($J_1$), 2NN Fe$_0$-Fe$_2$ ($J_2$), and 3NN Fe$_0$-Fe$_3$ ($J_3$), as seen in Fig \ref{4Jstructure}. Considering the magnetic exchange expression $-JS^2$ (FM for $J > 0$) for each pair of Fe$^{2+}$ $S = 2$ ions, we calculate the relative energies per formula unit of FePS$_3$ monolayer in the four magnetic structures.

\begin{equation}
	\begin{aligned}
		E_{\text{zigzag AF}} &= (-\frac{1}{2}J_1+J_2+\frac{3}{2}J_3)S^2\\
		E_{\text{FM}} &= (-\frac{3}{2}J_1-3J_2-\frac{3}{2}J_3)S^2\\
		E_{\text{N$\acute{\rm e}$el AF}} &= (+\frac{3}{2}J_1-3J_2+\frac{3}{2}J_3)S^2\\
		E_{\text{stripe AF}} &= (+\frac{1}{2}J_1+J_2-\frac{3}{2}J_3)S^2
	\end{aligned}
	\label{eq: 3}
\end{equation}

Using the relative total energies in Table \ref{tb3J} and applying Eq. \ref{eq: 3}, we determine the exchange parameters for FePS$_3$ monolayer as $J_1 = 3.13$ meV, $J_2 = -0.34$ meV, and $J_3 = -2.01$ meV. 
Our results indicate that the 1NN Fe$^{2+}$ ions, separated by 3.42 $\angstrom$, have a strong FM coupling, while the 2NN, at a distance of 5.93 $\angstrom$, exhibits a much weaker AF coupling. Intriguingly, the 3NN $J_3$ Fe$^{2+}$ ions, situated 6.84 $\angstrom$ apart—twice the 1NN distance—manifest a strong AF coupling. These results closely match the experimental ones from magnon bands measurement~\cite{Lan2016}. The interplay of 1NN FM and the long-range AF interactions of 3NN is crucial to the zigzag AF ground state observed in FePS$_3$ monolayer~\cite{Lee2016,zhang2021,Wang2016}.

\renewcommand\arraystretch{1.3}
\begin{table}[t]
	\centering
	\caption{Relative total energies $\Delta$\textit{E} (meV/f.u.), local spin and orbital moments ($\mu_{\rm B}$) and the derived three exchange parameters (meV) for FePS$_3$ and FePSe$_3$ monolayers by the GGA + SOC + $U$ calculations. 
}
	\begin{tabular}{c@{\hskip6mm}c@{\hskip6mm}c@{\hskip6mm}r@{\hskip6mm}r@{\hskip6mm}}
		\hline\hline
	Systems	& States  & $\Delta$\textit{E} & Fe$_{\rm spin}$  & Fe$_{\rm orb}$    \\ \hline
	\multirow{5}{*}{FePS$_3$}	    & zigzag AF  &  0       &  $\pm$3.51  &  $\pm$0.76      \\
		& FM      &  17.1           &  3.55       &  0.73          \\
	& 	N$\acute{e}$el AF  &  30.5         &  $\pm$3.49  &  $\pm$0.85      \\
		& stripe AF  &  36.7        &  $\pm$3.52  &  $\pm$0.78     \\
	&	\multicolumn{1}{c}{\textit{J}$_{1}$=3.13}    & \multicolumn{2}{c}{\textit{J}$_{2}$=--0.34} &    \multicolumn{1}{c}{\textit{J}$_{3}$=--2.01}    \\
		\hline
	\multirow{5}{*}{FePSe$_3$} &	zigzag AF  &  0       &  $\pm$3.41  &  $\pm$0.67      \\
		& FM      &  29.7           &  3.46       &  0.66          \\
		& 	N$\acute{e}$el AF  &  42.4         &  $\pm$3.38  &  $\pm$0.83      \\
		& stripe AF  &  43.7        &  $\pm$3.41  &  $\pm$0.69     \\
	&	\multicolumn{1}{c}{\textit{J}$_{1}$=3.53}    & \multicolumn{2}{c}{\textit{J}$_{2}$=--0.89} &    \multicolumn{1}{c}{\textit{J}$_{3}$=--2.47}    \\
		\hline\hline
	\end{tabular}
	\label{tb3J}
\end{table}

Here, we provide an insight into the 1NN FM coupling, the much weaker 2NN AF coupling, and the strong 3NN AF coupling, by examining the relevant hopping parameters through Wannier function analyses. Given the honeycomb lattice of the Fe$^{2+}$ magnetic ions and the edge-sharing FeS$_6$ octahedral network (see Fig .\ref{structure}(a)), here we adopt a local octahedral $XYZ$ coordinate system, where the $XYZ$ axes are directed from Fe to neighboring S (Se) ions. The eigenwave functions of the local octahedral structure under a trigonal crystal field can be described as

\begin{equation} \label{eq: 4}
	\begin{aligned}
		&\left|e_{g_{1,2}}^\sigma\right\rangle=\frac{1}{\sqrt{2}}(\left|3Z^{2}-R^2\right\rangle \pm \left| X^{2}-Y^{2}\right\rangle)\\
		&\left|a_{1g}\right\rangle=\frac{1}{\sqrt{3}}(\left|XY\right\rangle
		+\left|XZ\right\rangle+\left|YZ\right\rangle)\\
		&\left|L_{z\pm}\right\rangle=\frac{1}{\sqrt{3}}(\left|XY\right\rangle
		+e^{\pm \frac{i2 \pi}{3}}\left|XZ\right\rangle+e^{\pm  \frac{i4 \pi}{3}}\left|YZ\right\rangle) 
	\end{aligned}
\end{equation}

We select the projected Wannier orbitals by focusing exclusively on the Fe 3$d$-S 3$p$ hybrid orbitals near the Fermi level. This approach inherently encompasses both the indirect $d$-$p$-$d$ hoppings and the direct $d$-$d$ ones.
Figs. S2 and S3 in SM~\cite{SM} display the comparison between DFT and Wannier-interpolated band structures of the FePS$_3$ monolayer, showing that the chosen MLWFs accurately reproduce the $Ab$ $initio$ electronic states. With these Wannier functions, we can obtain the hopping parameters of different ions and orbitals.

\renewcommand\arraystretch{1.3}
\begin{table}[t]
	\centering
	\caption{The hopping parameters (meV) of 1NN Fe$_0$-Fe$_1$, 2NN Fe$_0$-Fe$_2$, 3NN Fe$_0$-Fe$_3$, and 4NN Fe$_0$-Fe$_4$ in FePS$_3$ monolayer.}
	\begin{tabular}{c@{\hskip3mm}l@{\hskip3mm}r@{\hskip3mm}r@{\hskip3mm}r@{\hskip3mm}r@{\hskip3mm}r@{\hskip3mm}} \hline\hline
		\multicolumn{2}{c}{\multirow{2}{*}{Hopping (\textit{t})}}  & \multicolumn{5}{c}{Fe$_0$} \\		      
		& & $3Z^2-R^2$ & $X^2-Y^2$ & $XY$ & $XZ$ & $YZ$ \\ \hline
		\multirow{5}{*}{Fe$_1$} & $3Z^2-R^2$ & --64 & 0 & 79 & 6 & 5 \\
		& $X^2-Y^2$ & 0 & --71 & 0 & 23 & --23 \\
		& $XY$ & 79 & 0 & --277 & 29 & 29 \\
		& $XZ$ & 6 & 23 & 29 & 74 & --43 \\
		& $YZ$ & 5 & --23 & 29 & --43 & 73 \\ \hline
		\multirow{5}{*}{Fe$_2$} & $3Z^2-R^2$ & 1 & 25 & 7 & --38 & 16 \\
		& $X^2-Y^2$ & 7 & 19 & 27 & --42 & --6 \\
		& $XY$ & --13 & 11 & 18 & --7 & --11 \\
		& $XZ$ & --17 & --54 & 15 & 0 & --7 \\
		& $YZ$ & 21 & 19 & --15 & 15 & 18 \\ \hline
		\multirow{5}{*}{Fe$_3$} & $3Z^2-R^2$ & 140 & 99 & 3 & --4 & --27 \\
		& $X^2-Y^2$ & 99 & 26 & 2 & --2 & 47 \\
		& $XY$ & 3 & 2 & 6 & --13 & --3 \\
		& $XZ$ & --4 & --2 & --13 & 6 & --2 \\
		& $YZ$ & --27 & 47 & --3 & --2 & --37 \\    \hline
		\multirow{5}{*}{Fe$_4$} & $3Z^2-R^2$ & 9 & 0 & 1 & --5 & 4 \\
		& $X^2-Y^2$ & 0 & --2 & --1 & --5 & 5 \\
		& $XY$ & 1 & --1 & 4 & --2 & --1 \\
		& $XZ$ & --5 & --5 & --2 & --3 & 9 \\
		& $YZ$ & 4 & 5 & --1 & 9 & 7 \\
		\hline\hline
	\end{tabular}
	\label{tbhopping}
\end{table}

\begin{figure}[t]
	\includegraphics[width=7.8cm]{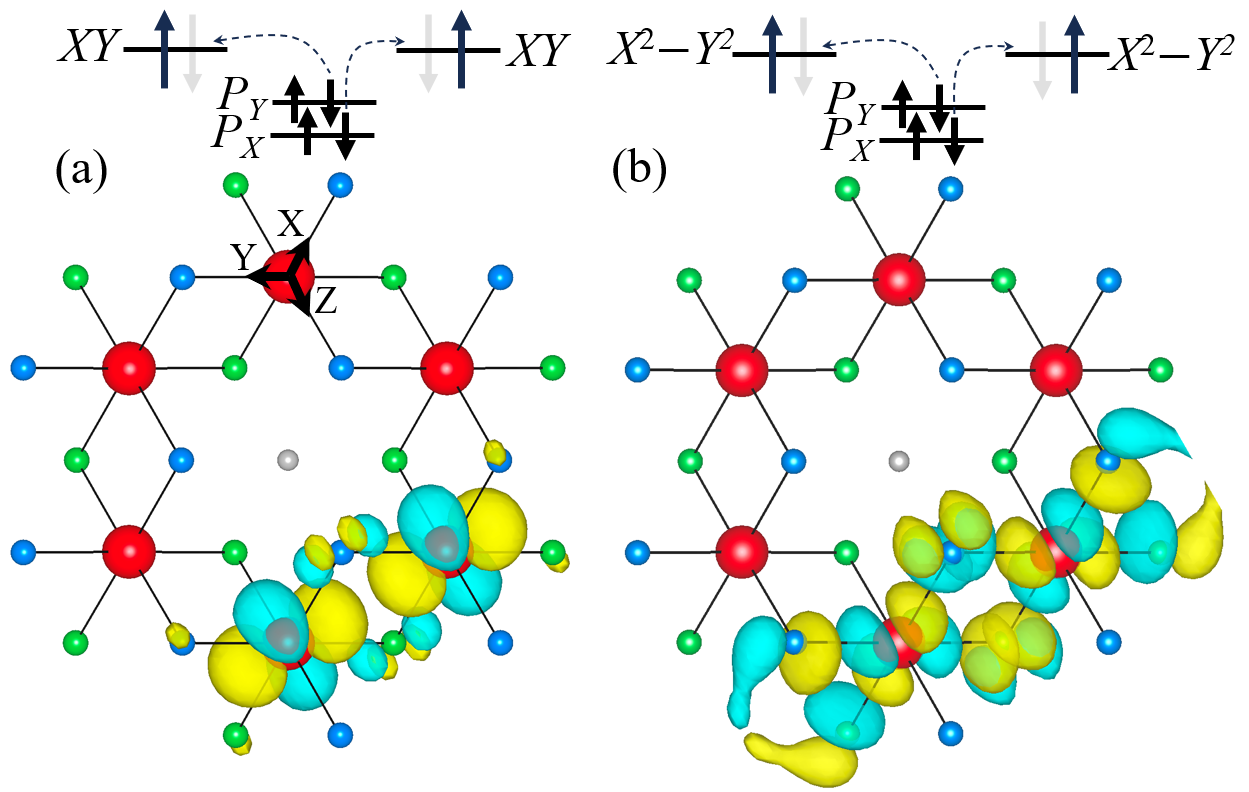}
	\centering
	\caption{Schematic plot of the 1NN FM superexchange and some associated hopping channels in FePS$_3$ monolayer represented by Wannier orbitals: the superexchange (a) via ($XY$)-($p_X$, $p_Y$)-($XY$) orbitals, and (b) via ($X^2-Y^2$)-($p_X$, $p_Y$)-($X^2-Y^2$).}
	\label{J1}
\end{figure}

We first examine the major hopping channels associated with the 1NN FM coupling in FePS$_3$ monolayer. In Table~\ref{tbhopping}, we list the hopping parameters and note that the $XY$ orbitals of Fe$_0$-Fe$_1$ ions show the largest hopping integral of 277 meV. Other hopping integrals are smaller than one-third of that value.
To understand why there is such a large hopping integral between the two $XY$ orbitals, we illustrate the real-space distribution of the $XY$-like MLWFs in Fig.~\ref{J1}(a).
In the edge-sharing octahedra, the $XY$ orbitals on adjacent Fe sites are directed toward each other. Considering the 1NN distance of 3.42 $\angstrom$, this leads to $dd\sigma$ hybridization.
Moreover, as seen in Fig.~\ref{J1}(a), besides the direct hopping integral, the $XY$ orbitals can interact through the indirect hoppings via sulfur 3$p$ orbitals, which is evident from the $pd\pi$ hybridization through the ($p_X$, $p_Y$) orbitals. In total, between adjacent $XY$ orbitals, there are not only direct hopping integrals but also indirect ones through the ligands, thus resulting in the largest hopping parameter. Then, we illustrate all hopping integrals exceeding 50 meV, as seen in Figs.~\ref{J1} and S4 in SM~\cite{SM}. We find that the direct $d$-$d$ hoppings are relatively small, while the indirect ones via ligands play a major role. For example, the superexchange channel involving ($X^2-Y^2$)-($p_X$, $p_Y$)-($X^2-Y^2$) primarily arises from $pd\sigma$ hybridization via the ($p_X$, $p_Y$) orbitals, as seen in Fig.~\ref{J1}(b). Considering virtual charge fluctuations, local Hund exchange, and the Pauli exclusion principle, those superexchange channels yield an FM coupling. Such near 90-degree 1NN superexchange channels leading to the FM $J_1$ have also been confirmed in other 2D FM semiconductors such as CrI$_3$~\cite{Lado2017,Xu2018,Wang_2016} and Cr$_2$Ge$_2$Te$_6$~\cite{Xu2018}.

As shown in Table~\ref{tbhopping}, the hopping parameters for 2NN Fe$_0$-Fe$_2$ are much smaller, aligning with our DFT calculations of small $J_2 = -0.34$ meV. Specifically, the most significant hopping integral for 2NN, arising from the indirect $d$-$p$-$d$ hopping, is 54 meV between the $XZ$ and $X^2-Y^2$ orbitals, as shown in Fig. S5 in SM~\cite{SM}. The hopping between these two orbitals is solely mediated by a $p_Y$ orbital, resulting in a much weaker hybridization. All other hopping integrals are smaller and thus all their contributions to the $J_2$ are quite limited.

By looking at the even smaller 4NN Fe$_0$-Fe$_4$ hoppings, their contributions to the 4NN superexchange should be negligibly weak. 
To obtain the $J_4$ exchange parameter, we additionally calculated a double stripe AF magnetic structure, as seen in Fig S6 in SM~\cite{SM}. We then find that $J_4$ is only $-$0.04 meV, and therefore, the tiny 4NN superexchange is of no more concern in this work.

\begin{figure}[t]
	\includegraphics[width=9cm]{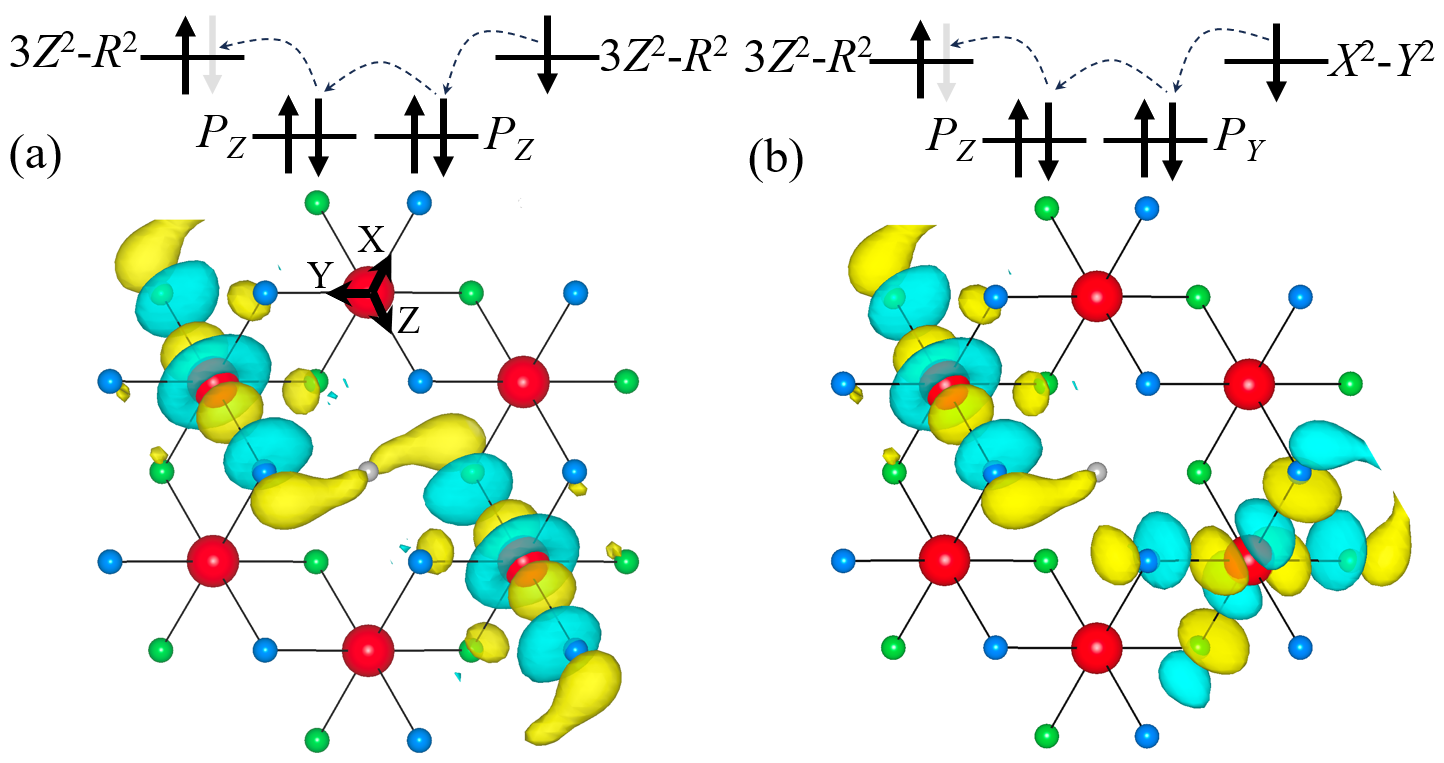}
	\centering
	\caption{Schematic plot of the 3NN AF superexchange and some associated hopping channels in FePS$_3$ monolayer represented by Wannier orbitals: the long-range superexchange (a) via ($3Z^2-R^2$)-($p_Z$)-($p_Z$)-($3Z^2-R^2$) orbitals, and (b) via  ($3Z^2-R^2$)-($p_Z$)-($p_Y$)-($X^2-Y^2$).}
	\label{J3}
\end{figure}

Notably, for 3NN Fe$_0$-Fe$_3$, the large distance of 6.84 $\angstrom$ is twice that of 1NN and thereby renders the direct $d$-$d$ hopping negligible. However, as seen in Table~\ref{tbhopping}, the hopping integral between the $3Z^2-R^2$ orbitals is 140 meV, and the integral between $3Z^2-R^2$ and $X^2-Y^2$ is 99 meV, both of which are significantly larger than most of the 1NN hopping parameters. Therefore, it is crucial to investigate how these $d$ orbitals via ligands facilitate such large hoppings and then contribute to the strong long-range superexchange interactions.
To illustrate the superexchange channels, we depict the real-space distribution of the $3Z^2-R^2$ and $X^2-Y^2$-like MLWFs in Fig.~\ref{J3}. Our results indicate that the $3Z^2-R^2$ orbitals of Fe$_0$ and Fe$_3$ ions engage in long-range superexchange via $p_Z$ orbitals of two sulfur ions. Each $3Z^2-R^2$ orbital forms a strong $pd\sigma$ hybridization with the adjacent sulfur $p_Z$ orbitals. Additionally, there is a hybridization between the two sulfur $p_Z$ orbitals mediated by the intermediate P atoms. This long-range hopping channel notably increases the hopping integral between the two $3Z^2-R^2$ orbitals. Similarly, the $3Z^2-R^2$ and $X^2-Y^2$ orbitals form a long-range hopping through two sulfur ions, facilitated by their respective $pd\sigma$ hybridizations with the $p_Z$ and $p_Y$ orbitals. The $p_Z$ and $p_Y$ orbitals on the same plane create a 90-degree head-to-head ($\frac{1}{2}$ $pp\sigma$$-$$\frac{1}{2}$ $pp\pi$) hybridization. Our results reveal that the effective hopping integral for the $3Z^2-R^2$ and $X^2-Y^2$ orbitals is smaller than the hopping integral between two $3Z^2-R^2$ orbitals, but these values are significantly larger than most of the 1NN hopping integrals.
Considering the $3d^{5\uparrow}L_{z+}^{1\downarrow}$ ground state, in which both the majority-spin $3Z^2-R^2$ and $X^2-Y^2$ orbitals are fully occupied by electrons, the above long-range superexchange channels consequently contribute to the strong $J_3$ AF coupling.

To summarize, our calculations confirm the zigzag AF ground state of FePS$_3$ monolayer, and find the competitive 1NN FM $J_1$ and 3NN AF $J_3$ but the much weaker 2NN AF $J_2$ and the negligibly weak 4NN AF $J_4$. In combination with Wannier function analyses and the derived hopping parameters, we find that the 1NN FM $J_1$ is primarily attributed to near 90-degree superexchange interactions associated with several channels. The strong 3NN AF $J_3$ arises from long-range superexchange interactions through two major channels: one is between two $3Z^2-R^2$ orbitals via the $p_Z$ orbitals of two sulfur ions (mediated by the intermediate P atoms) and their $pd\sigma$ hybridizations as shown in Fig.~\ref{J3}(a); the other involves $3Z^2-R^2$ and $X^2-Y^2$ orbitals via the S $p_Z$-S $p_Y$ channel and their respective $pd\sigma$ hybridizations as shown in Fig.~\ref{J3}(b). It is the competitive 1NN FM $J_1$ and 3NN AF $J_3$ that determine the zigzag AFM of FePS$_3$ monolayer.

\subsection*{D. The $T_ {\rm N}$ and strain effect} 

To estimate the  $T_ {\rm N}$ of FePS$_3$ monolayer, we assume a spin Hamiltonian and carry out PTMC simulations
\begin{equation} \label{eq: 5}
	H=-\sum_{k = 1, 2, 3}\sum_{i,j} \frac{J_{k}}{2} \mathbf{S}_{i} \cdot \mathbf{S}_{j}-\sum_{i} D\left(S_{i}^{z}\right)^{2}
\end{equation}   
The first term represents the isotropic Heisenberg exchange, and the sum runs over all Fe$^{2+}$ sites $i$ with $S$ = 2 in the spin-lattice, and $j$ runs over the $k$NN Fe$^{2+}$ sites of each $i$ with their respective magnetic couplings $J_k$ given as $J_1$ = 3.13 meV, $J_2$ = $-$0.34 meV and $J_3$ = $-$2.01 meV (FM when $J$ $>$ 0). The second term describes the MA with $S^z$ = 2 (easy perpendicular magnetization when $D$ $>$ 0). 
Given the giant SIA of 19.4 meV/Fe for the 3$d^{5\uparrow}L_{z+}^{1\downarrow}$ ground state, we obtain $D = 4.85$ meV. Employing these exchange parameters and the huge MA value, our PTMC simulations yield $T_{\rm N}$ = 119 K for FePS$_3$ monolayer as seen in Fig. \ref{strains}(a), and this agrees well with the experimental $T_{\rm N}$ = 118 K~\cite{Lee2016,zhang2021,Wang2016}.

\begin{figure}[t]
	\includegraphics[width=8cm]{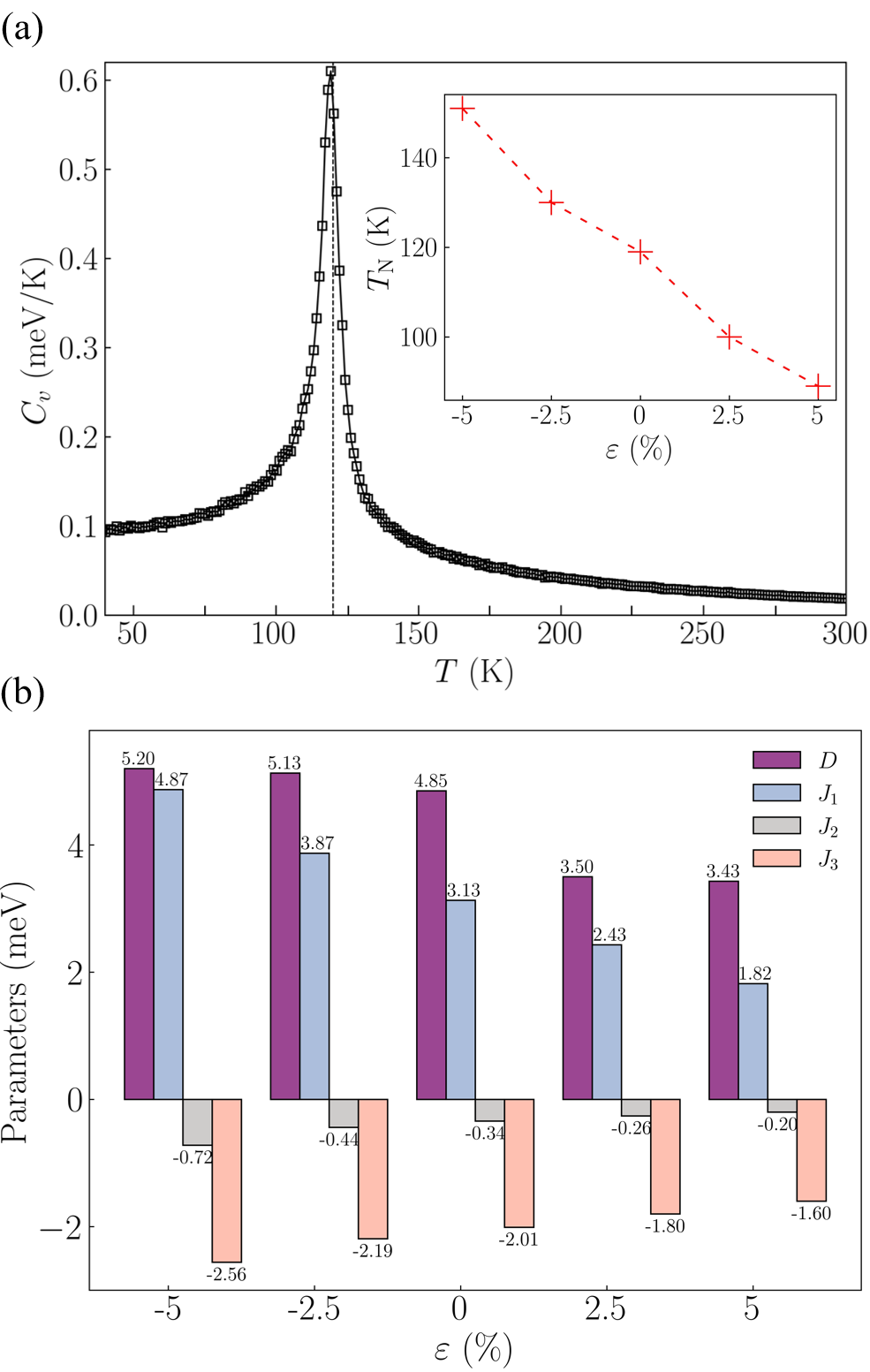}
	\centering
	\caption{(a) PTMC simulations of the magnetic specific heat of FePS$_3$ monolayer. The inset shows the increasing $T_{\rm N}$ under compressive strains. (b) the MA value $D$ and exchange parameters $J_1$, $J_2$, and $J_3$ (meV) of FePS$_3$ monolayer under the strains. 
}
	\label{strains}
\end{figure}

Strain is widely used to tune the properties of 2D materials~\cite{Cenker2022,Wang2020}.
Here we investigate a possible impact of a biaxial strain on FePS$_3$ monolayer.
Our calculations show that under strain, FePS$_3$ monolayer remains in the robust 3$d^{5\uparrow}L_{z+}^{1\downarrow}$ ground state. In particular, the compressive strains enhance the exchange parameters $J_i$ ($i$ = 1-3) and the MA value $D$, and thus boost $T_{\rm N}$ to 151 K under a $-$5\% compressive strain, as seen in Fig. ~\ref{strains}.

\subsection*{E. FePSe$_3$ monolayer: Ising zigzag AFM}

Bulk FePSe$_3$ is a zigzag AF semiconductor with $T_{\rm N}$ = 119 K~\cite{Wiedenmann1981}, and it has the same crystal structure as bulk FePS$_3$. FePSe$_3$ monolayer has not yet been experimentally synthesized to date. 
Here we also study the electronic structure and magnetism of FePSe$_3$ monolayer to check whether it is an Ising magnet too.

Our results indicate that FePSe$_3$ monolayer is in the same 3$d^{5\uparrow}L_{z+}^{1\downarrow}$ ground state as FePS$_3$ monolayer, by a direct comparison of different spin-orbital states, as seen in Fig.~\ref{states} and Fig. S7 in SM~\cite{SM}. It has the spin moment of 3.46 $\mu_{\rm B}$ in the FM state and the orbital moment of 0.66 $\mu_{\rm B}$ along the $z$-axis, see Table I. Both values are smaller than the corresponding ones of 3.55 $\mu_{\rm B}$ and 0.73 $\mu_{\rm B}$ in FePS$_3$ monolayer, and this is due to the stronger Fe-Se covalence reduction in FePSe$_3$. This is also in line with the experimental observations that bulk FePSe$_3$ has a smaller effective magnetic moment than bulk FePS$_3$ ~\cite{Joy1992,Wiedenmann1981}. Moreover, through our results in Fig. 4 and following the same procedures as in subsection III B, we estimate the SOC parameter $\zeta$ = 52 meV, and the trigonal crystal field splitting of 123 meV between the higher $a_{1g}$ singlet and lower $e_{g}^{\pi}$ doublet both out of the octahedral $t_{2g}$ triplet. The nice agreement with the typical Fe$^{2+}$ $\zeta$ parameter of 50-60 meV once again reflects the good accuracy of our calculations. The $a_{1g}$-$e_{g}^{\pi}$ splitting turns out to be much larger than the SOC parameter, and this enables us to restrict our discussion of the SOC effect within the half-filled minority-spin $e_{g}^{\pi}$ doublet. As FePSe$_3$ monolayer has a large orbital moment along the $z$-axis, which fixes via the SOC the spin orientation also along the $z$-axis, a tentative rotation of the spin moment into the $xy$ plane would cost a lot of the SOC energy. Indeed, our calculations find that the perpendicular MAE is 14.9 meV/Fe in FePSe$_3$ (see Fig. 4), which arises from the huge SIA associated with the 3$d^{5\uparrow}L_{z+}^{1\downarrow}$ ground state. Then, FePSe$_3$ monolayer is indeed an Ising magnet too.

\renewcommand\arraystretch{1.3}
\begin{table}[H]
	\centering
	\caption{The hopping parameters (meV) of 1NN Fe$_0$-Fe$_1$, 2NN Fe$_0$-Fe$_2$, 3NN Fe$_0$-Fe$_3$, and 4NN Fe$_0$-Fe$_4$ in FePSe$_3$ monolayer.}
	\begin{tabular}{c@{\hskip3mm}l@{\hskip3mm}r@{\hskip3mm}r@{\hskip3mm}r@{\hskip3mm}r@{\hskip3mm}r@{\hskip3mm}} \hline\hline
		\multicolumn{2}{c}{\multirow{2}{*}{Hopping (\textit{t})}}  & \multicolumn{5}{c}{Fe$_0$} \\		      
		& & $3Z^2-R^2$ & $X^2-Y^2$ & $XY$ & $XZ$ & $YZ$ \\ \hline
		\multirow{5}{*}{Fe$_1$} & $3Z^2-R^2$ & --62 & 0 & 105 & 7 & 5 \\
		& $X^2-Y^2$ & 0 & --45 & 0 & 33 & --32 \\
		& $XY$ & 105 & 0 & --229 & 23 & 23 \\
		& $XZ$ & 7 &  33 & 23 & 68 & --45 \\
		& $YZ$ & 5 & --32 & 23 & --45 & 68 \\ \hline
		\multirow{5}{*}{Fe$_2$} & $3Z^2-R^2$ & 5 & 18 & 3 & --31 & 13 \\
		& $X^2-Y^2$ & 15 & 25 & 34 & --36 & --12 \\
		& $XY$ & --17 & 6 & 17 & --9 & --23 \\
		& $XZ$ & --16 & --45 & 16 & --2 & --10 \\
		& $YZ$ & 28 & 19 & --19 & 16 & 17 \\ \hline
		\multirow{5}{*}{Fe$_3$} & $3Z^2-R^2$ & 128 & 96 & 2 & 3 & --30 \\
		& $X^2-Y^2$ & 96 & 18 & --5 & --5 & 52 \\
		& $XY$ & 2 & --5 & 5 & --12 & --1 \\
		& $XZ$ & 3 & --5 & --12 & 5 & 0 \\
		& $YZ$ & --30 & 52 & --1 & 0 & --40 \\    \hline
		\multirow{5}{*}{Fe$_4$} & $3Z^2-R^2$ & 12 & 4 & 2 & --6 & 5 \\
		& $X^2-Y^2$ & 4 & 2 & 2 & --8 & 7 \\
		& $XY$ & 2 & 2 & 5 & --2 & --2 \\
		& $XZ$ & --6 & --8 & --2 & --3 & 10 \\
		& $YZ$ & 5 & 7 & --2 & 10 & 5 \\
		\hline\hline
	\end{tabular}
	\label{tbhopping1}
\end{table}

\begin{figure}[H]
	\includegraphics[width=7.5cm]{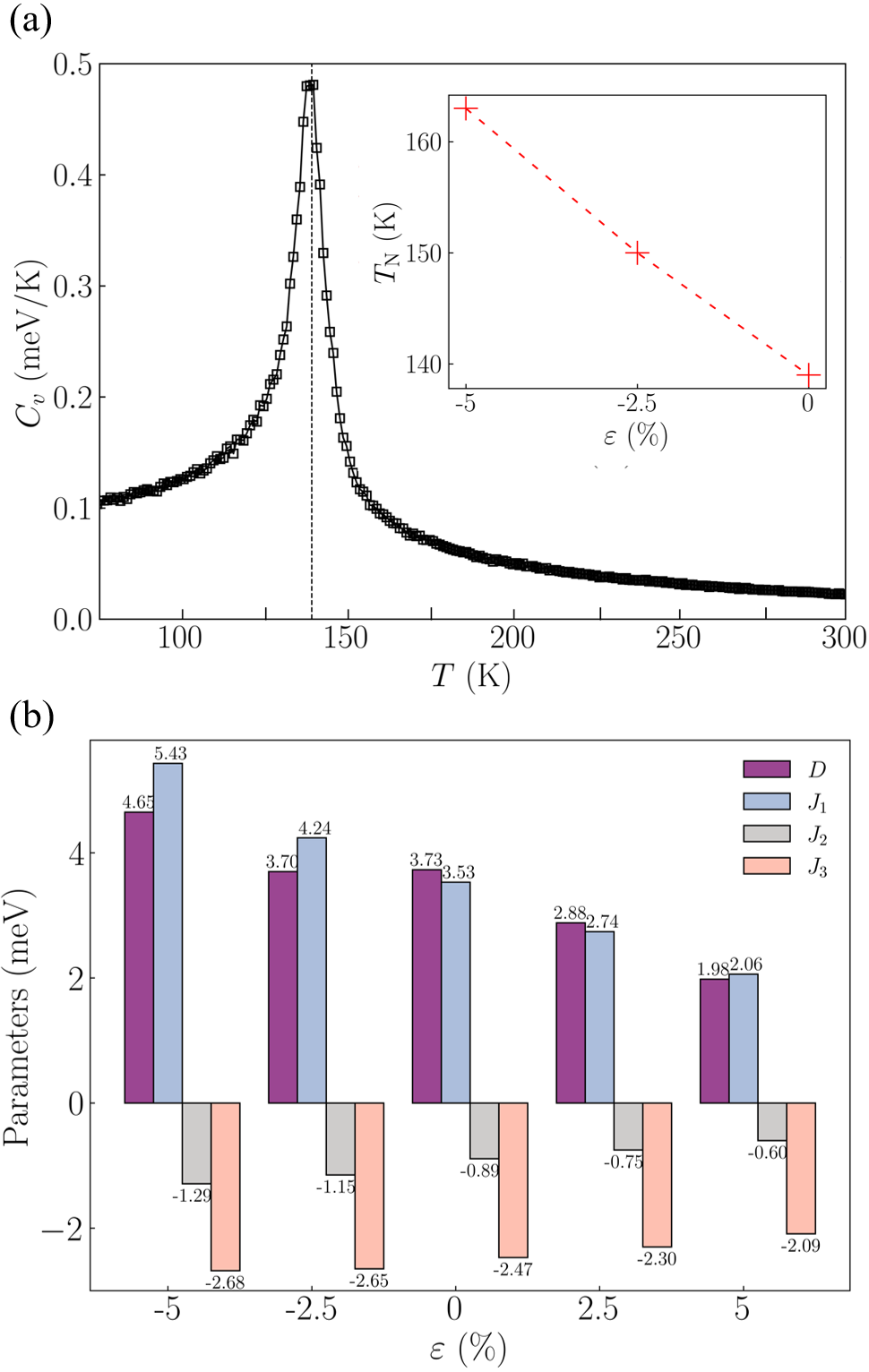}
	\centering
	\caption{(a) PTMC simulations of the magnetic specific heat of FePSe$_3$ monolayer. The inset shows the increasing $T_{\rm N}$ under compressive strains. (b) the MA value $D$ and exchange parameters $J_1$, $J_2$, and $J_3$ (meV) of FePSe$_3$ monolayer under the strains.     
	}
	\label{FePSe3_strains}
\end{figure}

By comparing four magnetic structures (Fig.~\ref{4Jstructure}) in our calculations, we find that FePSe$_3$ monolayer is a zigzag AF semiconductor with a band gap of 0.6 eV. The calculated exchange parameters are $J_1 = 3.53$ meV, $J_2 = -0.89$ meV, and $J_3 = -2.47$ meV, as seen in Table~\ref{tb3J}. 
We also calculate the hopping parameters for FePSe$_3$ monolayer using Wannier functions, as seen in Table ~\ref{tbhopping1}. These hopping parameters have the same tendency as those in FePS$_3$ monolayer (see Table ~\ref{tbhopping} for a comparison): the big 1NN and 3NN ones but much smaller 2NN and negligible 4NN, and they could help us to understand the competitive 1NN FM and 3NN AF but relatively weak 2NN AF and the negligible 4NN one, following the above discussion in Section III C. 
Note that the stronger Fe $3d$-Se $4p$ hybridizations, counteracting the increasing atomic distances in FePSe$_3$ normally with decreasing hopping integrals, give rise to quite similar hopping integrals $t$’s for FePSe$_3$ as in FePS$_3$, see Tables~\ref{tbhopping} and ~\ref{tbhopping1}. Then, together with the reduced Hubbard $U$, the superexchange parameters (roughly in the scale of $t^2$/$U$) seem larger in FePSe$_3$ than in FePS$_3$, as seen in Table~\ref{tb3J}. 
Then, using Eq.~\ref{eq: 5}, the aforementioned three exchange parameters and the MAE parameter for FePSe$_3$ monolayer, our PTMC simulations yield $T_{\rm N}$ = 140 K, and the $T_{\rm N}$ could be increased up to 163 K under the $-$5\% compressive strain, see Fig.~\ref{FePSe3_strains}.

Note that the $T_{\rm N}$ = 140 K for bare FePSe$_3$ monolayer seems overestimated, compared with the experimental $T_{\rm N}$ = 119 K for bulk FePSe$_3$~\cite{Wiedenmann1981}. This is associated with the likely overestimated exchange parameters due to the calculated smaller Hubbard $U$ = 2.7 eV from the constrained random phase approximation. When we choose the common value of $U$ = 4.0 eV in our GGA + SOC + $U$ calculations, the $T_{\rm N}$ is estimated to be 99 K for FePSe$_3$ monolayer (105 K for FePS$_3$ monolayer), see the results in Table S2 and Fig. S1 in SM~\cite{SM}. Note, however, that the Ising-type zigzag AF semiconducting ground state with the 3$d^{5\uparrow}L_{z+}^{1\downarrow}$ spin-orbital state remains unchanged at all.
The present results and prediction call for an experimental study on FePSe$_3$ monolayer.

\section*{IV. Conclusion}

To conclude, we study the electronic structure and magnetism of FePS$_3$ and FePSe$_3$ monolayers using density functional calculations, crystal field level diagrams, Wannier function analyses, and PTMC simulations. We find that both materials are in the robust Fe$^{2+}$ 3$d^{5\uparrow}L_{z+}^{1\downarrow}$ ground state with the formal $S_z$ = 2 and $L_z$ = 1. The large orbital moment produces a significant SIA and thus determines the Ising magnetism. The derived hopping parameters from Wannier functions help to explain the competitive 1NN FM and 3NN AF couplings but relatively weaker 2NN AF coupling, all of which determine the zigzag AFM. 
Our PTMC simulations well reproduce the experimental $T_{\rm N}$ = 118 K for FePS$_3$ monolayer and predict a close or even higher $T_{\rm N}$ for FePSe$_3$ monolayer, and moreover, their $T_{\rm N}$ could be enhanced under a compressive strain. This study provides an insight into the Ising-type zigzag AFM of FePS$_3$ and FePSe$_3$ monolayers.

\section*{Acknowledgements}
This work was supported by National Natural Science Foundation of China (Grants No. 12104307, No. 12174062 and No. 12241402), and by Innovation Program for Quantum Science and Technology.
\bibliography{FePS3_FePSe3.bib}

\begin{thebibliography}{51}%
\makeatletter
\providecommand \@ifxundefined [1]{%
 \@ifx{#1\undefined}
}%
\providecommand \@ifnum [1]{%
 \ifnum #1\expandafter \@firstoftwo
 \else \expandafter \@secondoftwo
 \fi
}%
\providecommand \@ifx [1]{%
 \ifx #1\expandafter \@firstoftwo
 \else \expandafter \@secondoftwo
 \fi
}%
\providecommand \natexlab [1]{#1}%
\providecommand \enquote  [1]{``#1''}%
\providecommand \bibnamefont  [1]{#1}%
\providecommand \bibfnamefont [1]{#1}%
\providecommand \citenamefont [1]{#1}%
\providecommand \href@noop [0]{\@secondoftwo}%
\providecommand \href [0]{\begingroup \@sanitize@url \@href}%
\providecommand \@href[1]{\@@startlink{#1}\@@href}%
\providecommand \@@href[1]{\endgroup#1\@@endlink}%
\providecommand \@sanitize@url [0]{\catcode `\\12\catcode `\$12\catcode
  `\&12\catcode `\#12\catcode `\^12\catcode `\_12\catcode `\%12\relax}%
\providecommand \@@startlink[1]{}%
\providecommand \@@endlink[0]{}%
\providecommand \url  [0]{\begingroup\@sanitize@url \@url }%
\providecommand \@url [1]{\endgroup\@href {#1}{\urlprefix }}%
\providecommand \urlprefix  [0]{URL }%
\providecommand \Eprint [0]{\href }%
\providecommand \doibase [0]{https://doi.org/}%
\providecommand \selectlanguage [0]{\@gobble}%
\providecommand \bibinfo  [0]{\@secondoftwo}%
\providecommand \bibfield  [0]{\@secondoftwo}%
\providecommand \translation [1]{[#1]}%
\providecommand \BibitemOpen [0]{}%
\providecommand \bibitemStop [0]{}%
\providecommand \bibitemNoStop [0]{.\EOS\space}%
\providecommand \EOS [0]{\spacefactor3000\relax}%
\providecommand \BibitemShut  [1]{\csname bibitem#1\endcsname}%
\let\auto@bib@innerbib\@empty
\bibitem [{\citenamefont {Novoselov}\ \emph {et~al.}(2004)\citenamefont
  {Novoselov}, \citenamefont {Geim}, \citenamefont {Morozov}, \citenamefont
  {Jiang}, \citenamefont {Zhang}, \citenamefont {Dubonos}, \citenamefont
  {Grigorieva},\ and\ \citenamefont {Firsov}}]{Novoselov2004}%
  \BibitemOpen
  \bibfield  {author} {\bibinfo {author} {\bibfnamefont {K.~S.}\ \bibnamefont
  {Novoselov}}, \bibinfo {author} {\bibfnamefont {A.~K.}\ \bibnamefont {Geim}},
  \bibinfo {author} {\bibfnamefont {S.~V.}\ \bibnamefont {Morozov}}, \bibinfo
  {author} {\bibfnamefont {D.}~\bibnamefont {Jiang}}, \bibinfo {author}
  {\bibfnamefont {Y.}~\bibnamefont {Zhang}}, \bibinfo {author} {\bibfnamefont
  {S.~V.}\ \bibnamefont {Dubonos}}, \bibinfo {author} {\bibfnamefont {I.~V.}\
  \bibnamefont {Grigorieva}},\ and\ \bibinfo {author} {\bibfnamefont {A.~A.}\
  \bibnamefont {Firsov}},\ }\bibfield  {title} {\bibinfo {title} {Electric
  field effect in atomically thin carbon films},\ }\href
  {https://www.science.org/doi/10.1126/science.1102896} {\bibfield  {journal}
  {\bibinfo  {journal} {Science}\ }\textbf {\bibinfo {volume} {306}},\ \bibinfo
  {pages} {666} (\bibinfo {year} {2004})}\BibitemShut {NoStop}%
\bibitem [{\citenamefont {Geim}\ and\ \citenamefont
  {Novoselov}(2007)}]{Geim2007}%
  \BibitemOpen
  \bibfield  {author} {\bibinfo {author} {\bibfnamefont {A.~K.}\ \bibnamefont
  {Geim}}\ and\ \bibinfo {author} {\bibfnamefont {K.~S.}\ \bibnamefont
  {Novoselov}},\ }\bibfield  {title} {\bibinfo {title} {The rise of graphene},\
  }\href {https://www.nature.com/articles/nmat1849} {\bibfield  {journal}
  {\bibinfo  {journal} {Nat. Mater.}\ }\textbf {\bibinfo {volume} {6}},\
  \bibinfo {pages} {183} (\bibinfo {year} {2007})}\BibitemShut {NoStop}%
\bibitem [{\citenamefont {Li}\ \emph {et~al.}(2019)\citenamefont {Li},
  \citenamefont {Jiang}, \citenamefont {Sivadas}, \citenamefont {Wang},
  \citenamefont {Xu}, \citenamefont {Weber}, \citenamefont {Goldberger},
  \citenamefont {Watanabe}, \citenamefont {Taniguchi}, \citenamefont {Fennie},
  \citenamefont {Mak},\ and\ \citenamefont {Shen}}]{Li2019}%
  \BibitemOpen
  \bibfield  {author} {\bibinfo {author} {\bibfnamefont {T.}~\bibnamefont
  {Li}}, \bibinfo {author} {\bibfnamefont {S.}~\bibnamefont {Jiang}}, \bibinfo
  {author} {\bibfnamefont {N.}~\bibnamefont {Sivadas}}, \bibinfo {author}
  {\bibfnamefont {Z.}~\bibnamefont {Wang}}, \bibinfo {author} {\bibfnamefont
  {Y.}~\bibnamefont {Xu}}, \bibinfo {author} {\bibfnamefont {D.}~\bibnamefont
  {Weber}}, \bibinfo {author} {\bibfnamefont {J.~E.}\ \bibnamefont
  {Goldberger}}, \bibinfo {author} {\bibfnamefont {K.}~\bibnamefont
  {Watanabe}}, \bibinfo {author} {\bibfnamefont {T.}~\bibnamefont {Taniguchi}},
  \bibinfo {author} {\bibfnamefont {C.~J.}\ \bibnamefont {Fennie}}, \bibinfo
  {author} {\bibfnamefont {K.~F.}\ \bibnamefont {Mak}},\ and\ \bibinfo {author}
  {\bibfnamefont {J.}~\bibnamefont {Shen}},\ }\bibfield  {title} {\bibinfo
  {title} {Pressure-controlled interlayer magnetism in atomically thin
  {CrI}$_{3}$},\ }\href {https://doi.org/10.1038/s41563-019-0506-1} {\bibfield
  {journal} {\bibinfo  {journal} {Nat. Mater.}\ }\textbf {\bibinfo {volume}
  {18}},\ \bibinfo {pages} {1303} (\bibinfo {year} {2019})}\BibitemShut
  {NoStop}%
\bibitem [{\citenamefont {Song}\ \emph {et~al.}(2019)\citenamefont {Song},
  \citenamefont {Fei}, \citenamefont {Yankowitz}, \citenamefont {Lin},
  \citenamefont {Jiang}, \citenamefont {Hwangbo}, \citenamefont {Zhang},
  \citenamefont {Sun}, \citenamefont {Taniguchi}, \citenamefont {Watanabe},
  \citenamefont {McGuire}, \citenamefont {Graf}, \citenamefont {Cao},
  \citenamefont {Chu}, \citenamefont {Cobden}, \citenamefont {Dean},
  \citenamefont {Xiao},\ and\ \citenamefont {Xu}}]{Song2019}%
  \BibitemOpen
  \bibfield  {author} {\bibinfo {author} {\bibfnamefont {T.}~\bibnamefont
  {Song}}, \bibinfo {author} {\bibfnamefont {Z.}~\bibnamefont {Fei}}, \bibinfo
  {author} {\bibfnamefont {M.}~\bibnamefont {Yankowitz}}, \bibinfo {author}
  {\bibfnamefont {Z.}~\bibnamefont {Lin}}, \bibinfo {author} {\bibfnamefont
  {Q.}~\bibnamefont {Jiang}}, \bibinfo {author} {\bibfnamefont
  {K.}~\bibnamefont {Hwangbo}}, \bibinfo {author} {\bibfnamefont
  {Q.}~\bibnamefont {Zhang}}, \bibinfo {author} {\bibfnamefont
  {B.}~\bibnamefont {Sun}}, \bibinfo {author} {\bibfnamefont {T.}~\bibnamefont
  {Taniguchi}}, \bibinfo {author} {\bibfnamefont {K.}~\bibnamefont {Watanabe}},
  \bibinfo {author} {\bibfnamefont {M.~A.}\ \bibnamefont {McGuire}}, \bibinfo
  {author} {\bibfnamefont {D.}~\bibnamefont {Graf}}, \bibinfo {author}
  {\bibfnamefont {T.}~\bibnamefont {Cao}}, \bibinfo {author} {\bibfnamefont
  {J.-H.}\ \bibnamefont {Chu}}, \bibinfo {author} {\bibfnamefont {D.~H.}\
  \bibnamefont {Cobden}}, \bibinfo {author} {\bibfnamefont {C.~R.}\
  \bibnamefont {Dean}}, \bibinfo {author} {\bibfnamefont {D.}~\bibnamefont
  {Xiao}},\ and\ \bibinfo {author} {\bibfnamefont {X.}~\bibnamefont {Xu}},\
  }\bibfield  {title} {\bibinfo {title} {Switching {2D} magnetic states via
  pressure tuning of layer stacking},\ }\href
  {https://doi.org/10.1038/s41563-019-0505-2} {\bibfield  {journal} {\bibinfo
  {journal} {Nat. Mater.}\ }\textbf {\bibinfo {volume} {18}},\ \bibinfo {pages}
  {1298} (\bibinfo {year} {2019})}\BibitemShut {NoStop}%
\bibitem [{\citenamefont {Li}\ \emph {et~al.}(2020)\citenamefont {Li},
  \citenamefont {Yang}, \citenamefont {Liu}, \citenamefont {Huang},
  \citenamefont {Wu}, \citenamefont {Zhang}, \citenamefont {Zhao},
  \citenamefont {Ma}, \citenamefont {Dang}, \citenamefont {Wei}, \citenamefont
  {Wang}, \citenamefont {Lin}, \citenamefont {Yan}, \citenamefont {Sun},
  \citenamefont {Li}, \citenamefont {Pan}, \citenamefont {Luo}, \citenamefont
  {Zhang}, \citenamefont {Liu}, \citenamefont {Huang}, \citenamefont {Duan},\
  and\ \citenamefont {Duan}}]{Li2020}%
  \BibitemOpen
  \bibfield  {author} {\bibinfo {author} {\bibfnamefont {J.}~\bibnamefont
  {Li}}, \bibinfo {author} {\bibfnamefont {X.}~\bibnamefont {Yang}}, \bibinfo
  {author} {\bibfnamefont {Y.}~\bibnamefont {Liu}}, \bibinfo {author}
  {\bibfnamefont {B.}~\bibnamefont {Huang}}, \bibinfo {author} {\bibfnamefont
  {R.}~\bibnamefont {Wu}}, \bibinfo {author} {\bibfnamefont {Z.}~\bibnamefont
  {Zhang}}, \bibinfo {author} {\bibfnamefont {B.}~\bibnamefont {Zhao}},
  \bibinfo {author} {\bibfnamefont {H.}~\bibnamefont {Ma}}, \bibinfo {author}
  {\bibfnamefont {W.}~\bibnamefont {Dang}}, \bibinfo {author} {\bibfnamefont
  {Z.}~\bibnamefont {Wei}}, \bibinfo {author} {\bibfnamefont {K.}~\bibnamefont
  {Wang}}, \bibinfo {author} {\bibfnamefont {Z.}~\bibnamefont {Lin}}, \bibinfo
  {author} {\bibfnamefont {X.}~\bibnamefont {Yan}}, \bibinfo {author}
  {\bibfnamefont {M.}~\bibnamefont {Sun}}, \bibinfo {author} {\bibfnamefont
  {B.}~\bibnamefont {Li}}, \bibinfo {author} {\bibfnamefont {X.}~\bibnamefont
  {Pan}}, \bibinfo {author} {\bibfnamefont {J.}~\bibnamefont {Luo}}, \bibinfo
  {author} {\bibfnamefont {G.}~\bibnamefont {Zhang}}, \bibinfo {author}
  {\bibfnamefont {Y.}~\bibnamefont {Liu}}, \bibinfo {author} {\bibfnamefont
  {Y.}~\bibnamefont {Huang}}, \bibinfo {author} {\bibfnamefont
  {X.}~\bibnamefont {Duan}},\ and\ \bibinfo {author} {\bibfnamefont
  {X.}~\bibnamefont {Duan}},\ }\bibfield  {title} {\bibinfo {title} {General
  synthesis of two-dimensional van der {Waals} heterostructure arrays},\ }\href
  {http://www.nature.com/articles/s41586-020-2098-y} {\bibfield  {journal}
  {\bibinfo  {journal} {Nature}\ }\textbf {\bibinfo {volume} {579}},\ \bibinfo
  {pages} {368} (\bibinfo {year} {2020})}\BibitemShut {NoStop}%
\bibitem [{\citenamefont {Burch}\ \emph {et~al.}(2018)\citenamefont {Burch},
  \citenamefont {Mandrus},\ and\ \citenamefont {Park}}]{Burch2018}%
  \BibitemOpen
  \bibfield  {author} {\bibinfo {author} {\bibfnamefont {K.~S.}\ \bibnamefont
  {Burch}}, \bibinfo {author} {\bibfnamefont {D.}~\bibnamefont {Mandrus}},\
  and\ \bibinfo {author} {\bibfnamefont {J.-G.}\ \bibnamefont {Park}},\
  }\bibfield  {title} {\bibinfo {title} {Magnetism in two-dimensional van der
  {Waals} materials},\ }\href
  {http://www.nature.com/articles/s41586-018-0631-z} {\bibfield  {journal}
  {\bibinfo  {journal} {Nature}\ }\textbf {\bibinfo {volume} {563}},\ \bibinfo
  {pages} {47} (\bibinfo {year} {2018})}\BibitemShut {NoStop}%
\bibitem [{\citenamefont {Song}\ \emph {et~al.}(2018)\citenamefont {Song},
  \citenamefont {Cai}, \citenamefont {Tu}, \citenamefont {Zhang}, \citenamefont
  {Huang}, \citenamefont {Wilson}, \citenamefont {Seyler}, \citenamefont {Zhu},
  \citenamefont {Taniguchi}, \citenamefont {Watanabe}, \citenamefont {McGuire},
  \citenamefont {Cobden}, \citenamefont {Xiao}, \citenamefont {Yao},\ and\
  \citenamefont {Xu}}]{Song2018}%
  \BibitemOpen
  \bibfield  {author} {\bibinfo {author} {\bibfnamefont {T.}~\bibnamefont
  {Song}}, \bibinfo {author} {\bibfnamefont {X.}~\bibnamefont {Cai}}, \bibinfo
  {author} {\bibfnamefont {M.~W.-Y.}\ \bibnamefont {Tu}}, \bibinfo {author}
  {\bibfnamefont {X.}~\bibnamefont {Zhang}}, \bibinfo {author} {\bibfnamefont
  {B.}~\bibnamefont {Huang}}, \bibinfo {author} {\bibfnamefont {N.~P.}\
  \bibnamefont {Wilson}}, \bibinfo {author} {\bibfnamefont {K.~L.}\
  \bibnamefont {Seyler}}, \bibinfo {author} {\bibfnamefont {L.}~\bibnamefont
  {Zhu}}, \bibinfo {author} {\bibfnamefont {T.}~\bibnamefont {Taniguchi}},
  \bibinfo {author} {\bibfnamefont {K.}~\bibnamefont {Watanabe}}, \bibinfo
  {author} {\bibfnamefont {M.~A.}\ \bibnamefont {McGuire}}, \bibinfo {author}
  {\bibfnamefont {D.~H.}\ \bibnamefont {Cobden}}, \bibinfo {author}
  {\bibfnamefont {D.}~\bibnamefont {Xiao}}, \bibinfo {author} {\bibfnamefont
  {W.}~\bibnamefont {Yao}},\ and\ \bibinfo {author} {\bibfnamefont
  {X.}~\bibnamefont {Xu}},\ }\bibfield  {title} {\bibinfo {title} {Giant
  tunneling magnetoresistance in spin-filter van der {Waals}
  heterostructures},\ }\href
  {http://www.sciencemag.org/lookup/doi/10.1126/science.aar4851} {\bibfield
  {journal} {\bibinfo  {journal} {Science}\ }\textbf {\bibinfo {volume}
  {360}},\ \bibinfo {pages} {1214} (\bibinfo {year} {2018})}\BibitemShut
  {NoStop}%
\bibitem [{\citenamefont {Huang}\ \emph {et~al.}(2017)\citenamefont {Huang},
  \citenamefont {Clark}, \citenamefont {Navarro-Moratalla}, \citenamefont
  {Klein}, \citenamefont {Cheng}, \citenamefont {Seyler}, \citenamefont
  {Zhong}, \citenamefont {Schmidgall}, \citenamefont {McGuire}, \citenamefont
  {Cobden}, \citenamefont {Yao}, \citenamefont {Xiao}, \citenamefont
  {Jarillo-Herrero},\ and\ \citenamefont {Xu}}]{Huang2017}%
  \BibitemOpen
  \bibfield  {author} {\bibinfo {author} {\bibfnamefont {B.}~\bibnamefont
  {Huang}}, \bibinfo {author} {\bibfnamefont {G.}~\bibnamefont {Clark}},
  \bibinfo {author} {\bibfnamefont {E.}~\bibnamefont {Navarro-Moratalla}},
  \bibinfo {author} {\bibfnamefont {D.~R.}\ \bibnamefont {Klein}}, \bibinfo
  {author} {\bibfnamefont {R.}~\bibnamefont {Cheng}}, \bibinfo {author}
  {\bibfnamefont {K.~L.}\ \bibnamefont {Seyler}}, \bibinfo {author}
  {\bibfnamefont {D.}~\bibnamefont {Zhong}}, \bibinfo {author} {\bibfnamefont
  {E.}~\bibnamefont {Schmidgall}}, \bibinfo {author} {\bibfnamefont {M.~A.}\
  \bibnamefont {McGuire}}, \bibinfo {author} {\bibfnamefont {D.~H.}\
  \bibnamefont {Cobden}}, \bibinfo {author} {\bibfnamefont {W.}~\bibnamefont
  {Yao}}, \bibinfo {author} {\bibfnamefont {D.}~\bibnamefont {Xiao}}, \bibinfo
  {author} {\bibfnamefont {P.}~\bibnamefont {Jarillo-Herrero}},\ and\ \bibinfo
  {author} {\bibfnamefont {X.}~\bibnamefont {Xu}},\ }\bibfield  {title}
  {\bibinfo {title} {Layer-dependent ferromagnetism in a van der {Waals}
  crystal down to the monolayer limit},\ }\href
  {https://doi.org/10.1038/nature22391} {\bibfield  {journal} {\bibinfo
  {journal} {Nature}\ }\textbf {\bibinfo {volume} {546}},\ \bibinfo {pages}
  {270} (\bibinfo {year} {2017})}\BibitemShut {NoStop}%
\bibitem [{\citenamefont {Gong}\ \emph {et~al.}(2017)\citenamefont {Gong},
  \citenamefont {Li}, \citenamefont {Li}, \citenamefont {Ji}, \citenamefont
  {Stern}, \citenamefont {Xia}, \citenamefont {Cao}, \citenamefont {Bao},
  \citenamefont {Wang}, \citenamefont {Wang}, \citenamefont {Qiu},
  \citenamefont {Cava}, \citenamefont {Louie}, \citenamefont {Xia},\ and\
  \citenamefont {Zhang}}]{Gong2017}%
  \BibitemOpen
  \bibfield  {author} {\bibinfo {author} {\bibfnamefont {C.}~\bibnamefont
  {Gong}}, \bibinfo {author} {\bibfnamefont {L.}~\bibnamefont {Li}}, \bibinfo
  {author} {\bibfnamefont {Z.}~\bibnamefont {Li}}, \bibinfo {author}
  {\bibfnamefont {H.}~\bibnamefont {Ji}}, \bibinfo {author} {\bibfnamefont
  {A.}~\bibnamefont {Stern}}, \bibinfo {author} {\bibfnamefont
  {Y.}~\bibnamefont {Xia}}, \bibinfo {author} {\bibfnamefont {T.}~\bibnamefont
  {Cao}}, \bibinfo {author} {\bibfnamefont {W.}~\bibnamefont {Bao}}, \bibinfo
  {author} {\bibfnamefont {C.}~\bibnamefont {Wang}}, \bibinfo {author}
  {\bibfnamefont {Y.}~\bibnamefont {Wang}}, \bibinfo {author} {\bibfnamefont
  {Z.~Q.}\ \bibnamefont {Qiu}}, \bibinfo {author} {\bibfnamefont {R.~J.}\
  \bibnamefont {Cava}}, \bibinfo {author} {\bibfnamefont {S.~G.}\ \bibnamefont
  {Louie}}, \bibinfo {author} {\bibfnamefont {J.}~\bibnamefont {Xia}},\ and\
  \bibinfo {author} {\bibfnamefont {X.}~\bibnamefont {Zhang}},\ }\bibfield
  {title} {\bibinfo {title} {Discovery of intrinsic ferromagnetism in
  two-dimensional van der {Waals} crystals},\ }\href
  {https://wwwnature.53yu.com/articles/nature22060} {\bibfield  {journal}
  {\bibinfo  {journal} {Nature}\ }\textbf {\bibinfo {volume} {546}},\ \bibinfo
  {pages} {265} (\bibinfo {year} {2017})}\BibitemShut {NoStop}%
\bibitem [{\citenamefont {Mermin}\ and\ \citenamefont
  {Wagner}(1966)}]{Mermin1966}%
  \BibitemOpen
  \bibfield  {author} {\bibinfo {author} {\bibfnamefont {N.~D.}\ \bibnamefont
  {Mermin}}\ and\ \bibinfo {author} {\bibfnamefont {H.}~\bibnamefont
  {Wagner}},\ }\bibfield  {title} {\bibinfo {title} {Absence of ferromagnetism
  or antiferromagnetism in one- or two-dimensional isotropic {Heisenberg}
  models},\ }\href {https://link.aps.org/doi/10.1103/PhysRevLett.17.1133}
  {\bibfield  {journal} {\bibinfo  {journal} {Phys. Rev. Lett.}\ }\textbf
  {\bibinfo {volume} {17}},\ \bibinfo {pages} {1133} (\bibinfo {year}
  {1966})}\BibitemShut {NoStop}%
\bibitem [{\citenamefont {Wang}\ \emph
  {et~al.}(2016{\natexlab{a}})\citenamefont {Wang}, \citenamefont {Fan},
  \citenamefont {Zhu},\ and\ \citenamefont {Wu}}]{Wang_2016}%
  \BibitemOpen
  \bibfield  {author} {\bibinfo {author} {\bibfnamefont {H.}~\bibnamefont
  {Wang}}, \bibinfo {author} {\bibfnamefont {F.}~\bibnamefont {Fan}}, \bibinfo
  {author} {\bibfnamefont {S.}~\bibnamefont {Zhu}},\ and\ \bibinfo {author}
  {\bibfnamefont {H.}~\bibnamefont {Wu}},\ }\bibfield  {title} {\bibinfo
  {title} {Doping enhanced ferromagnetism and induced half-metallicity in
  {CrI}$_{3}$ monolayer},\ }\href
  {https://dx.doi.org/10.1209/0295-5075/114/47001} {\bibfield  {journal}
  {\bibinfo  {journal} {Europhys. Lett.}\ }\textbf {\bibinfo {volume} {114}},\
  \bibinfo {pages} {47001} (\bibinfo {year} {2016}{\natexlab{a}})}\BibitemShut
  {NoStop}%
\bibitem [{\citenamefont {Lado}\ and\ \citenamefont
  {Fern{\'a}ndez-Rossier}(2017)}]{Lado2017}%
  \BibitemOpen
  \bibfield  {author} {\bibinfo {author} {\bibfnamefont {J.~L.}\ \bibnamefont
  {Lado}}\ and\ \bibinfo {author} {\bibfnamefont {J.}~\bibnamefont
  {Fern{\'a}ndez-Rossier}},\ }\bibfield  {title} {\bibinfo {title} {On the
  origin of magnetic anisotropy in two dimensional {CrI}$_{3}$},\ }\href
  {https://iopscience.iop.org/article/10.1088/2053-1583/aa75ed} {\bibfield
  {journal} {\bibinfo  {journal} {2D Mater.}\ }\textbf {\bibinfo {volume}
  {4}},\ \bibinfo {pages} {035002} (\bibinfo {year} {2017})}\BibitemShut
  {NoStop}%
\bibitem [{\citenamefont {Xu}\ \emph {et~al.}(2018)\citenamefont {Xu},
  \citenamefont {Feng}, \citenamefont {Xiang},\ and\ \citenamefont
  {Bellaiche}}]{Xu2018}%
  \BibitemOpen
  \bibfield  {author} {\bibinfo {author} {\bibfnamefont {C.}~\bibnamefont
  {Xu}}, \bibinfo {author} {\bibfnamefont {J.}~\bibnamefont {Feng}}, \bibinfo
  {author} {\bibfnamefont {H.}~\bibnamefont {Xiang}},\ and\ \bibinfo {author}
  {\bibfnamefont {L.}~\bibnamefont {Bellaiche}},\ }\bibfield  {title} {\bibinfo
  {title} {Interplay between {Kitaev} interaction and single ion anisotropy in
  ferromagnetic {CrI}$_{3}$ and {CrGeTe}$_{3}$ monolayers},\ }\href
  {https://doi.org/10.1038/s41524-018-0115-6} {\bibfield  {journal} {\bibinfo
  {journal} {npj Comput. Mater.}\ }\textbf {\bibinfo {volume} {4}},\ \bibinfo
  {pages} {57} (\bibinfo {year} {2018})}\BibitemShut {NoStop}%
\bibitem [{\citenamefont {Kim}\ \emph {et~al.}(2019)\citenamefont {Kim},
  \citenamefont {Kim}, \citenamefont {Ko}, \citenamefont {Seo}, \citenamefont
  {Kim}, \citenamefont {Jang}, \citenamefont {Kim}, \citenamefont {Kim},
  \citenamefont {Cheong},\ and\ \citenamefont {Park}}]{Kim2019}%
  \BibitemOpen
  \bibfield  {author} {\bibinfo {author} {\bibfnamefont {D.-H.}\ \bibnamefont
  {Kim}}, \bibinfo {author} {\bibfnamefont {K.}~\bibnamefont {Kim}}, \bibinfo
  {author} {\bibfnamefont {K.-T.}\ \bibnamefont {Ko}}, \bibinfo {author}
  {\bibfnamefont {J.}~\bibnamefont {Seo}}, \bibinfo {author} {\bibfnamefont
  {J.~S.}\ \bibnamefont {Kim}}, \bibinfo {author} {\bibfnamefont {T.-H.}\
  \bibnamefont {Jang}}, \bibinfo {author} {\bibfnamefont {Y.}~\bibnamefont
  {Kim}}, \bibinfo {author} {\bibfnamefont {J.-Y.}\ \bibnamefont {Kim}},
  \bibinfo {author} {\bibfnamefont {S.-W.}\ \bibnamefont {Cheong}},\ and\
  \bibinfo {author} {\bibfnamefont {J.-H.}\ \bibnamefont {Park}},\ }\bibfield
  {title} {\bibinfo {title} {Giant magnetic anisotropy induced by ligand ${LS}$
  coupling in layered {Cr} compounds},\ }\href
  {https://link.aps.org/doi/10.1103/PhysRevLett.122.207201} {\bibfield
  {journal} {\bibinfo  {journal} {Phys. Rev. Lett.}\ }\textbf {\bibinfo
  {volume} {122}},\ \bibinfo {pages} {207201} (\bibinfo {year}
  {2019})}\BibitemShut {NoStop}%
\bibitem [{\citenamefont {Yang}\ \emph {et~al.}(2020)\citenamefont {Yang},
  \citenamefont {Fan}, \citenamefont {Wang}, \citenamefont {Khomskii},\ and\
  \citenamefont {Wu}}]{Yang2020}%
  \BibitemOpen
  \bibfield  {author} {\bibinfo {author} {\bibfnamefont {K.}~\bibnamefont
  {Yang}}, \bibinfo {author} {\bibfnamefont {F.}~\bibnamefont {Fan}}, \bibinfo
  {author} {\bibfnamefont {H.}~\bibnamefont {Wang}}, \bibinfo {author}
  {\bibfnamefont {D.~I.}\ \bibnamefont {Khomskii}},\ and\ \bibinfo {author}
  {\bibfnamefont {H.}~\bibnamefont {Wu}},\ }\bibfield  {title} {\bibinfo
  {title} {{VI}$_{3}$: A two-dimensional {Ising} ferromagnet},\ }\href
  {https://link.aps.org/doi/10.1103/PhysRevB.101.100402} {\bibfield  {journal}
  {\bibinfo  {journal} {Phys. Rev. B}\ }\textbf {\bibinfo {volume} {101}},\
  \bibinfo {pages} {100402} (\bibinfo {year} {2020})}\BibitemShut {NoStop}%
\bibitem [{\citenamefont {Zhao}\ \emph {et~al.}(2021)\citenamefont {Zhao},
  \citenamefont {Liu}, \citenamefont {Hu}, \citenamefont {Jia}, \citenamefont
  {Cui}, \citenamefont {Wu}, \citenamefont {Whangbo},\ and\ \citenamefont
  {Ren}}]{Zhao2021}%
  \BibitemOpen
  \bibfield  {author} {\bibinfo {author} {\bibfnamefont {G.-D.}\ \bibnamefont
  {Zhao}}, \bibinfo {author} {\bibfnamefont {X.}~\bibnamefont {Liu}}, \bibinfo
  {author} {\bibfnamefont {T.}~\bibnamefont {Hu}}, \bibinfo {author}
  {\bibfnamefont {F.}~\bibnamefont {Jia}}, \bibinfo {author} {\bibfnamefont
  {Y.}~\bibnamefont {Cui}}, \bibinfo {author} {\bibfnamefont {W.}~\bibnamefont
  {Wu}}, \bibinfo {author} {\bibfnamefont {M.-H.}\ \bibnamefont {Whangbo}},\
  and\ \bibinfo {author} {\bibfnamefont {W.}~\bibnamefont {Ren}},\ }\bibfield
  {title} {\bibinfo {title} {Difference in magnetic anisotropy of the
  ferromagnetic monolayers {VI}$_{3}$ and {CrI}$_{3}$},\ }\href
  {https://link.aps.org/doi/10.1103/PhysRevB.103.014438} {\bibfield  {journal}
  {\bibinfo  {journal} {Phys. Rev. B}\ }\textbf {\bibinfo {volume} {103}},\
  \bibinfo {pages} {014438} (\bibinfo {year} {2021})}\BibitemShut {NoStop}%
\bibitem [{\citenamefont {De~Vita}\ \emph {et~al.}(2022)\citenamefont
  {De~Vita}, \citenamefont {Nguyen}, \citenamefont {Sant}, \citenamefont
  {Pierantozzi}, \citenamefont {Amoroso}, \citenamefont {Bigi}, \citenamefont
  {Polewczyk}, \citenamefont {Vinai}, \citenamefont {Nguyen}, \citenamefont
  {Kong}, \citenamefont {Fujii}, \citenamefont {Vobornik}, \citenamefont
  {Brookes}, \citenamefont {Rossi}, \citenamefont {Cava}, \citenamefont
  {Mazzola}, \citenamefont {Yamauchi}, \citenamefont {Picozzi},\ and\
  \citenamefont {Panaccione}}]{DeVita2022}%
  \BibitemOpen
  \bibfield  {author} {\bibinfo {author} {\bibfnamefont {A.}~\bibnamefont
  {De~Vita}}, \bibinfo {author} {\bibfnamefont {T.~T.~P.}\ \bibnamefont
  {Nguyen}}, \bibinfo {author} {\bibfnamefont {R.}~\bibnamefont {Sant}},
  \bibinfo {author} {\bibfnamefont {G.~M.}\ \bibnamefont {Pierantozzi}},
  \bibinfo {author} {\bibfnamefont {D.}~\bibnamefont {Amoroso}}, \bibinfo
  {author} {\bibfnamefont {C.}~\bibnamefont {Bigi}}, \bibinfo {author}
  {\bibfnamefont {V.}~\bibnamefont {Polewczyk}}, \bibinfo {author}
  {\bibfnamefont {G.}~\bibnamefont {Vinai}}, \bibinfo {author} {\bibfnamefont
  {L.~T.}\ \bibnamefont {Nguyen}}, \bibinfo {author} {\bibfnamefont
  {T.}~\bibnamefont {Kong}}, \bibinfo {author} {\bibfnamefont {J.}~\bibnamefont
  {Fujii}}, \bibinfo {author} {\bibfnamefont {I.}~\bibnamefont {Vobornik}},
  \bibinfo {author} {\bibfnamefont {N.~B.}\ \bibnamefont {Brookes}}, \bibinfo
  {author} {\bibfnamefont {G.}~\bibnamefont {Rossi}}, \bibinfo {author}
  {\bibfnamefont {R.~J.}\ \bibnamefont {Cava}}, \bibinfo {author}
  {\bibfnamefont {F.}~\bibnamefont {Mazzola}}, \bibinfo {author} {\bibfnamefont
  {K.}~\bibnamefont {Yamauchi}}, \bibinfo {author} {\bibfnamefont
  {S.}~\bibnamefont {Picozzi}},\ and\ \bibinfo {author} {\bibfnamefont
  {G.}~\bibnamefont {Panaccione}},\ }\bibfield  {title} {\bibinfo {title}
  {Influence of orbital character on the ground state electronic properties in
  the van der {Waals} transition metal iodides {VI}$_{3}$ and {CrI}$_{3}$},\
  }\href {https://pubs.acs.org/doi/10.1021/acs.nanolett.2c01922} {\bibfield
  {journal} {\bibinfo  {journal} {Nano Lett.}\ }\textbf {\bibinfo {volume}
  {22}},\ \bibinfo {pages} {7034} (\bibinfo {year} {2022})}\BibitemShut
  {NoStop}%
\bibitem [{\citenamefont {Lin}\ \emph {et~al.}(2021)\citenamefont {Lin},
  \citenamefont {Huang}, \citenamefont {Hwangbo}, \citenamefont {Jiang},
  \citenamefont {Zhang}, \citenamefont {Liu}, \citenamefont {Fei},
  \citenamefont {Lv}, \citenamefont {Millis}, \citenamefont {McGuire},
  \citenamefont {Xiao}, \citenamefont {Chu},\ and\ \citenamefont
  {Xu}}]{Lin2021}%
  \BibitemOpen
  \bibfield  {author} {\bibinfo {author} {\bibfnamefont {Z.}~\bibnamefont
  {Lin}}, \bibinfo {author} {\bibfnamefont {B.}~\bibnamefont {Huang}}, \bibinfo
  {author} {\bibfnamefont {K.}~\bibnamefont {Hwangbo}}, \bibinfo {author}
  {\bibfnamefont {Q.}~\bibnamefont {Jiang}}, \bibinfo {author} {\bibfnamefont
  {Q.}~\bibnamefont {Zhang}}, \bibinfo {author} {\bibfnamefont
  {Z.}~\bibnamefont {Liu}}, \bibinfo {author} {\bibfnamefont {Z.}~\bibnamefont
  {Fei}}, \bibinfo {author} {\bibfnamefont {H.}~\bibnamefont {Lv}}, \bibinfo
  {author} {\bibfnamefont {A.}~\bibnamefont {Millis}}, \bibinfo {author}
  {\bibfnamefont {M.}~\bibnamefont {McGuire}}, \bibinfo {author} {\bibfnamefont
  {D.}~\bibnamefont {Xiao}}, \bibinfo {author} {\bibfnamefont {J.-H.}\
  \bibnamefont {Chu}},\ and\ \bibinfo {author} {\bibfnamefont {X.}~\bibnamefont
  {Xu}},\ }\bibfield  {title} {\bibinfo {title} {Magnetism and its structural
  coupling effects in {2D} {Ising} ferromagnetic insulator {VI}$_{3}$},\ }\href
  {https://pubs.acs.org/doi/10.1021/acs.nanolett.1c03027} {\bibfield  {journal}
  {\bibinfo  {journal} {Nano Lett.}\ }\textbf {\bibinfo {volume} {21}},\
  \bibinfo {pages} {9180} (\bibinfo {year} {2021})}\BibitemShut {NoStop}%
\bibitem [{\citenamefont {Hovancik}\ \emph {et~al.}(2023)\citenamefont
  {Hovancik}, \citenamefont {Pospisil}, \citenamefont {Carva}, \citenamefont
  {Sechovsky},\ and\ \citenamefont {Piamonteze}}]{Hovancik2023}%
  \BibitemOpen
  \bibfield  {author} {\bibinfo {author} {\bibfnamefont {D.}~\bibnamefont
  {Hovancik}}, \bibinfo {author} {\bibfnamefont {J.}~\bibnamefont {Pospisil}},
  \bibinfo {author} {\bibfnamefont {K.}~\bibnamefont {Carva}}, \bibinfo
  {author} {\bibfnamefont {V.}~\bibnamefont {Sechovsky}},\ and\ \bibinfo
  {author} {\bibfnamefont {C.}~\bibnamefont {Piamonteze}},\ }\bibfield  {title}
  {\bibinfo {title} {Large orbital magnetic moment in {VI}$_{3}$},\ }\href
  {https://pubs.acs.org/doi/10.1021/acs.nanolett.2c04045} {\bibfield  {journal}
  {\bibinfo  {journal} {Nano Lett.}\ }\textbf {\bibinfo {volume} {23}},\
  \bibinfo {pages} {1175} (\bibinfo {year} {2023})}\BibitemShut {NoStop}%
\bibitem [{\citenamefont {Gibertini}\ \emph {et~al.}(2019)\citenamefont
  {Gibertini}, \citenamefont {Koperski}, \citenamefont {Morpurgo},\ and\
  \citenamefont {Novoselov}}]{Gibertini2019}%
  \BibitemOpen
  \bibfield  {author} {\bibinfo {author} {\bibfnamefont {M.}~\bibnamefont
  {Gibertini}}, \bibinfo {author} {\bibfnamefont {M.}~\bibnamefont {Koperski}},
  \bibinfo {author} {\bibfnamefont {A.~F.}\ \bibnamefont {Morpurgo}},\ and\
  \bibinfo {author} {\bibfnamefont {K.~S.}\ \bibnamefont {Novoselov}},\
  }\bibfield  {title} {\bibinfo {title} {Magnetic {2D} materials and
  heterostructures},\ }\href
  {https://www.nature.com/articles/s41565-019-0438-6} {\bibfield  {journal}
  {\bibinfo  {journal} {Nat. Nanotechnol.}\ }\textbf {\bibinfo {volume} {14}},\
  \bibinfo {pages} {408} (\bibinfo {year} {2019})}\BibitemShut {NoStop}%
\bibitem [{\citenamefont {Mak}\ \emph {et~al.}(2019)\citenamefont {Mak},
  \citenamefont {Shan},\ and\ \citenamefont {Ralph}}]{Mak2019}%
  \BibitemOpen
  \bibfield  {author} {\bibinfo {author} {\bibfnamefont {K.~F.}\ \bibnamefont
  {Mak}}, \bibinfo {author} {\bibfnamefont {J.}~\bibnamefont {Shan}},\ and\
  \bibinfo {author} {\bibfnamefont {D.~C.}\ \bibnamefont {Ralph}},\ }\bibfield
  {title} {\bibinfo {title} {Probing and controlling magnetic states in {2D}
  layered magnetic materials},\ }\href
  {https://www.nature.com/articles/s42254-019-0110-y} {\bibfield  {journal}
  {\bibinfo  {journal} {Nat. Rev. Phys.}\ }\textbf {\bibinfo {volume} {1}},\
  \bibinfo {pages} {646} (\bibinfo {year} {2019})}\BibitemShut {NoStop}%
\bibitem [{\citenamefont {Gong}\ and\ \citenamefont {Zhang}(2019)}]{Gong2019}%
  \BibitemOpen
  \bibfield  {author} {\bibinfo {author} {\bibfnamefont {C.}~\bibnamefont
  {Gong}}\ and\ \bibinfo {author} {\bibfnamefont {X.}~\bibnamefont {Zhang}},\
  }\bibfield  {title} {\bibinfo {title} {Two-dimensional magnetic crystals and
  emergent heterostructure devices},\ }\href
  {https://www.science.org/doi/10.1126/science.aav4450} {\bibfield  {journal}
  {\bibinfo  {journal} {Science}\ }\textbf {\bibinfo {volume} {363}},\ \bibinfo
  {pages} {706} (\bibinfo {year} {2019})}\BibitemShut {NoStop}%
\bibitem [{\citenamefont {Wiedenmann}\ \emph {et~al.}(1981)\citenamefont
  {Wiedenmann}, \citenamefont {Rossat-Mignod}, \citenamefont {Louisy},
  \citenamefont {Brec},\ and\ \citenamefont {Rouxel}}]{Wiedenmann1981}%
  \BibitemOpen
  \bibfield  {author} {\bibinfo {author} {\bibfnamefont {A.}~\bibnamefont
  {Wiedenmann}}, \bibinfo {author} {\bibfnamefont {J.}~\bibnamefont
  {Rossat-Mignod}}, \bibinfo {author} {\bibfnamefont {A.}~\bibnamefont
  {Louisy}}, \bibinfo {author} {\bibfnamefont {R.}~\bibnamefont {Brec}},\ and\
  \bibinfo {author} {\bibfnamefont {J.}~\bibnamefont {Rouxel}},\ }\bibfield
  {title} {\bibinfo {title} {Neutron diffraction study of the layered compounds
  {MnPSe}$_{3}$ and {FePSe}$_{3}$},\ }\href
  {https://www.sciencedirect.com/science/article/pii/0038109881902532}
  {\bibfield  {journal} {\bibinfo  {journal} {Solid State Commun.}\ }\textbf
  {\bibinfo {volume} {40}},\ \bibinfo {pages} {1067} (\bibinfo {year}
  {1981})}\BibitemShut {NoStop}%
\bibitem [{\citenamefont {Lee}\ \emph {et~al.}(2016)\citenamefont {Lee},
  \citenamefont {Lee}, \citenamefont {Ryoo}, \citenamefont {Kang},
  \citenamefont {Kim}, \citenamefont {Kim}, \citenamefont {Park}, \citenamefont
  {Park},\ and\ \citenamefont {Cheong}}]{Lee2016}%
  \BibitemOpen
  \bibfield  {author} {\bibinfo {author} {\bibfnamefont {J.-U.}\ \bibnamefont
  {Lee}}, \bibinfo {author} {\bibfnamefont {S.}~\bibnamefont {Lee}}, \bibinfo
  {author} {\bibfnamefont {J.~H.}\ \bibnamefont {Ryoo}}, \bibinfo {author}
  {\bibfnamefont {S.}~\bibnamefont {Kang}}, \bibinfo {author} {\bibfnamefont
  {T.~Y.}\ \bibnamefont {Kim}}, \bibinfo {author} {\bibfnamefont
  {P.}~\bibnamefont {Kim}}, \bibinfo {author} {\bibfnamefont {C.-H.}\
  \bibnamefont {Park}}, \bibinfo {author} {\bibfnamefont {J.-G.}\ \bibnamefont
  {Park}},\ and\ \bibinfo {author} {\bibfnamefont {H.}~\bibnamefont {Cheong}},\
  }\bibfield  {title} {\bibinfo {title} {Ising-type magnetic ordering in
  atomically thin {FePS}$_{3}$},\ }\href
  {https://pubs.acs.org/doi/10.1021/acs.nanolett.6b03052} {\bibfield  {journal}
  {\bibinfo  {journal} {Nano Lett.}\ }\textbf {\bibinfo {volume} {16}},\
  \bibinfo {pages} {7433} (\bibinfo {year} {2016})}\BibitemShut {NoStop}%
\bibitem [{\citenamefont {Coak}\ \emph {et~al.}(2019)\citenamefont {Coak},
  \citenamefont {Jarvis}, \citenamefont {Hamidov}, \citenamefont {Haines},
  \citenamefont {Alireza}, \citenamefont {Liu}, \citenamefont {Son},
  \citenamefont {Hwang}, \citenamefont {Lampronti}, \citenamefont
  {Daisenberger}, \citenamefont {Nahai-Williamson}, \citenamefont {Wildes},
  \citenamefont {Saxena},\ and\ \citenamefont {Park}}]{Coak2019}%
  \BibitemOpen
  \bibfield  {author} {\bibinfo {author} {\bibfnamefont {M.~J.}\ \bibnamefont
  {Coak}}, \bibinfo {author} {\bibfnamefont {D.~M.}\ \bibnamefont {Jarvis}},
  \bibinfo {author} {\bibfnamefont {H.}~\bibnamefont {Hamidov}}, \bibinfo
  {author} {\bibfnamefont {C.~R.~S.}\ \bibnamefont {Haines}}, \bibinfo {author}
  {\bibfnamefont {P.~L.}\ \bibnamefont {Alireza}}, \bibinfo {author}
  {\bibfnamefont {C.}~\bibnamefont {Liu}}, \bibinfo {author} {\bibfnamefont
  {S.}~\bibnamefont {Son}}, \bibinfo {author} {\bibfnamefont {I.}~\bibnamefont
  {Hwang}}, \bibinfo {author} {\bibfnamefont {G.~I.}\ \bibnamefont
  {Lampronti}}, \bibinfo {author} {\bibfnamefont {D.}~\bibnamefont
  {Daisenberger}}, \bibinfo {author} {\bibfnamefont {P.}~\bibnamefont
  {Nahai-Williamson}}, \bibinfo {author} {\bibfnamefont {A.~R.}\ \bibnamefont
  {Wildes}}, \bibinfo {author} {\bibfnamefont {S.~S.}\ \bibnamefont {Saxena}},\
  and\ \bibinfo {author} {\bibfnamefont {J.-G.}\ \bibnamefont {Park}},\
  }\bibfield  {title} {\bibinfo {title} {Tuning dimensionality in
  van-der-{Waals} antiferromagnetic {Mott} insulators {$TM$PS}$_{3}$},\ }\href
  {https://dx.doi.org/10.1088/1361-648X/ab5be8} {\bibfield  {journal} {\bibinfo
   {journal} {J. Phys.: Condens. Matter}\ }\textbf {\bibinfo {volume} {32}},\
  \bibinfo {pages} {124003} (\bibinfo {year} {2019})}\BibitemShut {NoStop}%
\bibitem [{\citenamefont {Zheng}\ \emph {et~al.}(2019)\citenamefont {Zheng},
  \citenamefont {Jiang}, \citenamefont {Xue}, \citenamefont {Dai},\ and\
  \citenamefont {Feng}}]{Zheng2019}%
  \BibitemOpen
  \bibfield  {author} {\bibinfo {author} {\bibfnamefont {Y.}~\bibnamefont
  {Zheng}}, \bibinfo {author} {\bibfnamefont {X.-x.}\ \bibnamefont {Jiang}},
  \bibinfo {author} {\bibfnamefont {X.-x.}\ \bibnamefont {Xue}}, \bibinfo
  {author} {\bibfnamefont {J.}~\bibnamefont {Dai}},\ and\ \bibinfo {author}
  {\bibfnamefont {Y.}~\bibnamefont {Feng}},\ }\bibfield  {title} {\bibinfo
  {title} {{$Ab$} $initio$ study of pressure-driven phase transition in
  {FePS}$_{3}$ and {FePSe}$_{3}$},\ }\href
  {https://link.aps.org/doi/10.1103/PhysRevB.100.174102} {\bibfield  {journal}
  {\bibinfo  {journal} {Phys. Rev. B}\ }\textbf {\bibinfo {volume} {100}},\
  \bibinfo {pages} {174102} (\bibinfo {year} {2019})}\BibitemShut {NoStop}%
\bibitem [{\citenamefont {Joy}\ and\ \citenamefont
  {Vasudevan}(1992)}]{Joy1992}%
  \BibitemOpen
  \bibfield  {author} {\bibinfo {author} {\bibfnamefont {P.~A.}\ \bibnamefont
  {Joy}}\ and\ \bibinfo {author} {\bibfnamefont {S.}~\bibnamefont
  {Vasudevan}},\ }\bibfield  {title} {\bibinfo {title} {Magnetism in the
  layered transition-metal thiophosphates {$M$PS}$_{3}$ ({$M$} = {Mn}, {Fe},
  and {Ni})},\ }\href {https://link.aps.org/doi/10.1103/PhysRevB.46.5425}
  {\bibfield  {journal} {\bibinfo  {journal} {Phys. Rev. B}\ }\textbf {\bibinfo
  {volume} {46}},\ \bibinfo {pages} {5425} (\bibinfo {year}
  {1992})}\BibitemShut {NoStop}%
\bibitem [{\citenamefont {Zhang}\ \emph
  {et~al.}(2021{\natexlab{a}})\citenamefont {Zhang}, \citenamefont {Hwangbo},
  \citenamefont {Wang}, \citenamefont {Jiang}, \citenamefont {Chu},
  \citenamefont {Wen}, \citenamefont {Xiao},\ and\ \citenamefont
  {Xu}}]{zhang2021}%
  \BibitemOpen
  \bibfield  {author} {\bibinfo {author} {\bibfnamefont {Q.}~\bibnamefont
  {Zhang}}, \bibinfo {author} {\bibfnamefont {K.}~\bibnamefont {Hwangbo}},
  \bibinfo {author} {\bibfnamefont {C.}~\bibnamefont {Wang}}, \bibinfo {author}
  {\bibfnamefont {Q.}~\bibnamefont {Jiang}}, \bibinfo {author} {\bibfnamefont
  {J.-H.}\ \bibnamefont {Chu}}, \bibinfo {author} {\bibfnamefont
  {H.}~\bibnamefont {Wen}}, \bibinfo {author} {\bibfnamefont {D.}~\bibnamefont
  {Xiao}},\ and\ \bibinfo {author} {\bibfnamefont {X.}~\bibnamefont {Xu}},\
  }\bibfield  {title} {\bibinfo {title} {Observation of giant optical linear
  dichroism in a zigzag antiferromagnet {FePS$_3$}},\ }\href
  {https://doi.org/10.1021/acs.nanolett.1c02188} {\bibfield  {journal}
  {\bibinfo  {journal} {Nano Lett.}\ }\textbf {\bibinfo {volume} {21}},\
  \bibinfo {pages} {6938} (\bibinfo {year} {2021}{\natexlab{a}})}\BibitemShut
  {NoStop}%
\bibitem [{\citenamefont {Wang}\ \emph
  {et~al.}(2016{\natexlab{b}})\citenamefont {Wang}, \citenamefont {Du},
  \citenamefont {Liu}, \citenamefont {Hu}, \citenamefont {Zhang}, \citenamefont
  {Zhang}, \citenamefont {Owen}, \citenamefont {Lu}, \citenamefont {Gan},
  \citenamefont {Sengupta}, \citenamefont {Kloc},\ and\ \citenamefont
  {Xiong}}]{Wang2016}%
  \BibitemOpen
  \bibfield  {author} {\bibinfo {author} {\bibfnamefont {X.}~\bibnamefont
  {Wang}}, \bibinfo {author} {\bibfnamefont {K.}~\bibnamefont {Du}}, \bibinfo
  {author} {\bibfnamefont {Y.~Y.~F.}\ \bibnamefont {Liu}}, \bibinfo {author}
  {\bibfnamefont {P.}~\bibnamefont {Hu}}, \bibinfo {author} {\bibfnamefont
  {J.}~\bibnamefont {Zhang}}, \bibinfo {author} {\bibfnamefont
  {Q.}~\bibnamefont {Zhang}}, \bibinfo {author} {\bibfnamefont {M.~H.~S.}\
  \bibnamefont {Owen}}, \bibinfo {author} {\bibfnamefont {X.}~\bibnamefont
  {Lu}}, \bibinfo {author} {\bibfnamefont {C.~K.}\ \bibnamefont {Gan}},
  \bibinfo {author} {\bibfnamefont {P.}~\bibnamefont {Sengupta}}, \bibinfo
  {author} {\bibfnamefont {C.}~\bibnamefont {Kloc}},\ and\ \bibinfo {author}
  {\bibfnamefont {Q.}~\bibnamefont {Xiong}},\ }\bibfield  {title} {\bibinfo
  {title} {Raman spectroscopy of atomically thin two-dimensional magnetic iron
  phosphorus trisulfide ({FePS$_3$}) crystals},\ }\href
  {https://dx.doi.org/10.1088/2053-1583/3/3/031009} {\bibfield  {journal}
  {\bibinfo  {journal} {2D Mater.}\ }\textbf {\bibinfo {volume} {3}},\ \bibinfo
  {pages} {031009} (\bibinfo {year} {2016}{\natexlab{b}})}\BibitemShut
  {NoStop}%
\bibitem [{\citenamefont {Chittari}\ \emph {et~al.}(2016)\citenamefont
  {Chittari}, \citenamefont {Park}, \citenamefont {Lee}, \citenamefont {Han},
  \citenamefont {MacDonald}, \citenamefont {Hwang},\ and\ \citenamefont
  {Jung}}]{Chittari2016}%
  \BibitemOpen
  \bibfield  {author} {\bibinfo {author} {\bibfnamefont {B.~L.}\ \bibnamefont
  {Chittari}}, \bibinfo {author} {\bibfnamefont {Y.}~\bibnamefont {Park}},
  \bibinfo {author} {\bibfnamefont {D.}~\bibnamefont {Lee}}, \bibinfo {author}
  {\bibfnamefont {M.}~\bibnamefont {Han}}, \bibinfo {author} {\bibfnamefont
  {A.~H.}\ \bibnamefont {MacDonald}}, \bibinfo {author} {\bibfnamefont
  {E.}~\bibnamefont {Hwang}},\ and\ \bibinfo {author} {\bibfnamefont
  {J.}~\bibnamefont {Jung}},\ }\bibfield  {title} {\bibinfo {title} {Electronic
  and magnetic properties of single-layer {$M$}{P}{$X$}$_{3}$ metal phosphorous
  trichalcogenides},\ }\href
  {https://link.aps.org/doi/10.1103/PhysRevB.94.184428} {\bibfield  {journal}
  {\bibinfo  {journal} {Phys. Rev. B}\ }\textbf {\bibinfo {volume} {94}},\
  \bibinfo {pages} {184428} (\bibinfo {year} {2016})}\BibitemShut {NoStop}%
\bibitem [{\citenamefont {Olsen}(2021)}]{Olsen2021}%
  \BibitemOpen
  \bibfield  {author} {\bibinfo {author} {\bibfnamefont {T.}~\bibnamefont
  {Olsen}},\ }\bibfield  {title} {\bibinfo {title} {Magnetic anisotropy and
  exchange interactions of two-dimensional {FePS}$_{3}$, {NiPS}$_{3}$ and
  {MnPS}$_{3}$ from first principles calculations},\ }\href
  {https://dx.doi.org/10.1088/1361-6463/ac000e} {\bibfield  {journal} {\bibinfo
   {journal} {J. Phys. D: Appl. Phys.}\ }\textbf {\bibinfo {volume} {54}},\
  \bibinfo {pages} {314001} (\bibinfo {year} {2021})}\BibitemShut {NoStop}%
\bibitem [{\citenamefont {Zhang}\ \emph
  {et~al.}(2021{\natexlab{b}})\citenamefont {Zhang}, \citenamefont {Nie},
  \citenamefont {Wang}, \citenamefont {Xia},\ and\ \citenamefont
  {Guo}}]{Zhang_2021}%
  \BibitemOpen
  \bibfield  {author} {\bibinfo {author} {\bibfnamefont {J.-m.}\ \bibnamefont
  {Zhang}}, \bibinfo {author} {\bibfnamefont {Y.-z.}\ \bibnamefont {Nie}},
  \bibinfo {author} {\bibfnamefont {X.-g.}\ \bibnamefont {Wang}}, \bibinfo
  {author} {\bibfnamefont {Q.-l.}\ \bibnamefont {Xia}},\ and\ \bibinfo {author}
  {\bibfnamefont {G.-h.}\ \bibnamefont {Guo}},\ }\bibfield  {title} {\bibinfo
  {title} {Strain modulation of magnetic properties of monolayer and bilayer
  {FePS}$_{3}$ antiferromagnet},\ }\href
  {https://www.sciencedirect.com/science/article/pii/S0304885320326548?via%3Dihub}
  {\bibfield  {journal} {\bibinfo  {journal} {J. Magn. Magn. Mater.}\ }\textbf
  {\bibinfo {volume} {525}},\ \bibinfo {pages} {167687} (\bibinfo {year}
  {2021}{\natexlab{b}})}\BibitemShut {NoStop}%
\bibitem [{\citenamefont {Kim}\ and\ \citenamefont {Park}(2021)}]{Kim2021}%
  \BibitemOpen
  \bibfield  {author} {\bibinfo {author} {\bibfnamefont {T.~Y.}\ \bibnamefont
  {Kim}}\ and\ \bibinfo {author} {\bibfnamefont {C.-H.}\ \bibnamefont {Park}},\
  }\bibfield  {title} {\bibinfo {title} {Magnetic anisotropy and magnetic
  ordering of transition-metal phosphorus trisulfides},\ }\href
  {https://pubs.acs.org/doi/10.1021/acs.nanolett.1c03992} {\bibfield  {journal}
  {\bibinfo  {journal} {Nano Lett.}\ }\textbf {\bibinfo {volume} {21}},\
  \bibinfo {pages} {10114} (\bibinfo {year} {2021})}\BibitemShut {NoStop}%
\bibitem [{\citenamefont {Amirabbasi}\ and\ \citenamefont
  {Kratzer}(2023)}]{Amirabbasi2023}%
  \BibitemOpen
  \bibfield  {author} {\bibinfo {author} {\bibfnamefont {M.}~\bibnamefont
  {Amirabbasi}}\ and\ \bibinfo {author} {\bibfnamefont {P.}~\bibnamefont
  {Kratzer}},\ }\bibfield  {title} {\bibinfo {title} {Orbital and magnetic
  ordering in single-layer {FePS}$_{3}$: A {DFT} + {$U$} study},\ }\href
  {https://doi.org/10.1103/PhysRevB.107.024401} {\bibfield  {journal} {\bibinfo
   {journal} {Phys. Rev. B}\ }\textbf {\bibinfo {volume} {107}},\ \bibinfo
  {pages} {024401} (\bibinfo {year} {2023})}\BibitemShut {NoStop}%
\bibitem [{\citenamefont {Li}\ \emph {et~al.}(2024)\citenamefont {Li},
  \citenamefont {Li}, \citenamefont {Feng}, \citenamefont {Ni}, \citenamefont
  {Guo},\ and\ \citenamefont {Xiang}}]{Li2024}%
  \BibitemOpen
  \bibfield  {author} {\bibinfo {author} {\bibfnamefont {P.}~\bibnamefont
  {Li}}, \bibinfo {author} {\bibfnamefont {X.}~\bibnamefont {Li}}, \bibinfo
  {author} {\bibfnamefont {J.}~\bibnamefont {Feng}}, \bibinfo {author}
  {\bibfnamefont {J.}~\bibnamefont {Ni}}, \bibinfo {author} {\bibfnamefont
  {Z.-X.}\ \bibnamefont {Guo}},\ and\ \bibinfo {author} {\bibfnamefont
  {H.}~\bibnamefont {Xiang}},\ }\bibfield  {title} {\bibinfo {title} {Origin of
  zigzag antiferromagnetic order in {$X$PS}$_{3}$ {(X = Fe, Ni)} monolayers},\
  }\href {https://doi.org/10.1103/PhysRevB.109.214418} {\bibfield  {journal}
  {\bibinfo  {journal} {Phys. Rev. B}\ }\textbf {\bibinfo {volume} {109}},\
  \bibinfo {pages} {214418} (\bibinfo {year} {2024})}\BibitemShut {NoStop}%
\bibitem [{\citenamefont {Kresse}\ and\ \citenamefont
  {Hafner}(1993)}]{Kresse1993}%
  \BibitemOpen
  \bibfield  {author} {\bibinfo {author} {\bibfnamefont {G.}~\bibnamefont
  {Kresse}}\ and\ \bibinfo {author} {\bibfnamefont {J.}~\bibnamefont
  {Hafner}},\ }\bibfield  {title} {\bibinfo {title} {{$Ab$} $initio$ molecular
  dynamics for liquid metals},\ }\href
  {https://link.aps.org/doi/10.1103/PhysRevB.47.558} {\bibfield  {journal}
  {\bibinfo  {journal} {Phys. Rev. B}\ }\textbf {\bibinfo {volume} {47}},\
  \bibinfo {pages} {558} (\bibinfo {year} {1993})}\BibitemShut {NoStop}%
\bibitem [{\citenamefont {Perdew}\ \emph {et~al.}(1996)\citenamefont {Perdew},
  \citenamefont {Burke},\ and\ \citenamefont {Ernzerhof}}]{Perdew1996}%
  \BibitemOpen
  \bibfield  {author} {\bibinfo {author} {\bibfnamefont {J.~P.}\ \bibnamefont
  {Perdew}}, \bibinfo {author} {\bibfnamefont {K.}~\bibnamefont {Burke}},\ and\
  \bibinfo {author} {\bibfnamefont {M.}~\bibnamefont {Ernzerhof}},\ }\bibfield
  {title} {\bibinfo {title} {Generalized gradient approximation made simple},\
  }\href {https://link.aps.org/doi/10.1103/PhysRevLett.77.3865} {\bibfield
  {journal} {\bibinfo  {journal} {Phys. Rev. Lett.}\ }\textbf {\bibinfo
  {volume} {77}},\ \bibinfo {pages} {3865} (\bibinfo {year}
  {1996})}\BibitemShut {NoStop}%
\bibitem [{\citenamefont {Ouvrard}\ \emph {et~al.}(1985)\citenamefont
  {Ouvrard}, \citenamefont {Brec},\ and\ \citenamefont {Rouxel}}]{Ouvrard1985}%
  \BibitemOpen
  \bibfield  {author} {\bibinfo {author} {\bibfnamefont {G.}~\bibnamefont
  {Ouvrard}}, \bibinfo {author} {\bibfnamefont {R.}~\bibnamefont {Brec}},\ and\
  \bibinfo {author} {\bibfnamefont {J.}~\bibnamefont {Rouxel}},\ }\bibfield
  {title} {\bibinfo {title} {Structural determination of some {$M$PS}$_{3}$
  layered phases ({$M$} = {Mn}, {Fe}, {Co}, {Ni} and {Cd})},\ }\href
  {https://www.sciencedirect.com/science/article/pii/0025540885900923}
  {\bibfield  {journal} {\bibinfo  {journal} {Mater. Res. Bull.}\ }\textbf
  {\bibinfo {volume} {20}},\ \bibinfo {pages} {1181} (\bibinfo {year}
  {1985})}\BibitemShut {NoStop}%
\bibitem [{SM()}]{SM}%
  \BibitemOpen
  \href@noop {} {}\bibinfo {note} {See Supplemental Material at
  http://link.aps.org/supplemental/ for the convergence tests, the tested U = 4
  eV results for FePS$_3$ and FePSe$_3$ monolayers, the 1NN and 2NN hopping
  channels, and the DOS results of FePSe$_3$ monolayer.}\BibitemShut {Stop}%
\bibitem [{\citenamefont {Anisimov}\ \emph {et~al.}(1997)\citenamefont
  {Anisimov}, \citenamefont {Aryasetiawan},\ and\ \citenamefont
  {Lichtenstein}}]{Anisimov1997}%
  \BibitemOpen
  \bibfield  {author} {\bibinfo {author} {\bibfnamefont {V.~I.}\ \bibnamefont
  {Anisimov}}, \bibinfo {author} {\bibfnamefont {F.}~\bibnamefont
  {Aryasetiawan}},\ and\ \bibinfo {author} {\bibfnamefont {A.~I.}\ \bibnamefont
  {Lichtenstein}},\ }\bibfield  {title} {\bibinfo {title} {First-principles
  calculations of the electronic structure and spectra of strongly correlated
  systems: the {LDA} + ${U}$ method},\ }\href
  {https://dx.doi.org/10.1088/0953-8984/9/4/002} {\bibfield  {journal}
  {\bibinfo  {journal} {J. Phys.: Condens. Matter}\ }\textbf {\bibinfo {volume}
  {9}},\ \bibinfo {pages} {767} (\bibinfo {year} {1997})}\BibitemShut {NoStop}%
\bibitem [{\citenamefont {Vaugier}\ \emph {et~al.}(2012)\citenamefont
  {Vaugier}, \citenamefont {Jiang},\ and\ \citenamefont {Biermann}}]{CRPA}%
  \BibitemOpen
  \bibfield  {author} {\bibinfo {author} {\bibfnamefont {L.}~\bibnamefont
  {Vaugier}}, \bibinfo {author} {\bibfnamefont {H.}~\bibnamefont {Jiang}},\
  and\ \bibinfo {author} {\bibfnamefont {S.}~\bibnamefont {Biermann}},\
  }\bibfield  {title} {\bibinfo {title} {Hubbard {$U$} and {Hund} exchange
  {$J$} in transition metal oxides: Screening versus localization trends from
  constrained random phase approximation},\ }\href
  {https://link.aps.org/doi/10.1103/PhysRevB.86.165105} {\bibfield  {journal}
  {\bibinfo  {journal} {Phys. Rev. B}\ }\textbf {\bibinfo {volume} {86}},\
  \bibinfo {pages} {165105} (\bibinfo {year} {2012})}\BibitemShut {NoStop}%
\bibitem [{\citenamefont {Khomskii}(2014)}]{Khomskii_2014}%
  \BibitemOpen
  \bibfield  {author} {\bibinfo {author} {\bibfnamefont {D.~I.}\ \bibnamefont
  {Khomskii}},\ }\href@noop {} {\emph {\bibinfo {title} {Transition Metal
  Compounds}}}\ (\bibinfo  {publisher} {Cambridge University Press},\ \bibinfo
  {year} {2014})\BibitemShut {NoStop}%
\bibitem [{\citenamefont {Allen}\ and\ \citenamefont
  {Watson}(2014)}]{Allen2014}%
  \BibitemOpen
  \bibfield  {author} {\bibinfo {author} {\bibfnamefont {J.~P.}\ \bibnamefont
  {Allen}}\ and\ \bibinfo {author} {\bibfnamefont {G.~W.}\ \bibnamefont
  {Watson}},\ }\bibfield  {title} {\bibinfo {title} {Occupation matrix control
  of d- and f-electron localisations using {DFT} + ${U}$},\ }\href
  {https://pubs.rsc.org/en/content/articlelanding/2014/CP/C4CP01083C}
  {\bibfield  {journal} {\bibinfo  {journal} {Phys. Chem. Chem. Phys.}\
  }\textbf {\bibinfo {volume} {16}},\ \bibinfo {pages} {21016} (\bibinfo {year}
  {2014})}\BibitemShut {NoStop}%
\bibitem [{\citenamefont {Mostofi}\ \emph {et~al.}(2008)\citenamefont
  {Mostofi}, \citenamefont {Yates}, \citenamefont {Lee}, \citenamefont {Souza},
  \citenamefont {Vanderbilt},\ and\ \citenamefont {Marzari}}]{Mostofi2008}%
  \BibitemOpen
  \bibfield  {author} {\bibinfo {author} {\bibfnamefont {A.~A.}\ \bibnamefont
  {Mostofi}}, \bibinfo {author} {\bibfnamefont {J.~R.}\ \bibnamefont {Yates}},
  \bibinfo {author} {\bibfnamefont {Y.-S.}\ \bibnamefont {Lee}}, \bibinfo
  {author} {\bibfnamefont {I.}~\bibnamefont {Souza}}, \bibinfo {author}
  {\bibfnamefont {D.}~\bibnamefont {Vanderbilt}},\ and\ \bibinfo {author}
  {\bibfnamefont {N.}~\bibnamefont {Marzari}},\ }\bibfield  {title} {\bibinfo
  {title} {wannier90: A tool for obtaining maximally-localised {Wannier}
  functions},\ }\href
  {https://www.sciencedirect.com/science/article/pii/S0010465507004936}
  {\bibfield  {journal} {\bibinfo  {journal} {Comput. Phys. Commun.}\ }\textbf
  {\bibinfo {volume} {178}},\ \bibinfo {pages} {685} (\bibinfo {year}
  {2008})}\BibitemShut {NoStop}%
\bibitem [{\citenamefont {Marzari}\ \emph {et~al.}(2012)\citenamefont
  {Marzari}, \citenamefont {Mostofi}, \citenamefont {Yates}, \citenamefont
  {Souza},\ and\ \citenamefont {Vanderbilt}}]{Nicola2012}%
  \BibitemOpen
  \bibfield  {author} {\bibinfo {author} {\bibfnamefont {N.}~\bibnamefont
  {Marzari}}, \bibinfo {author} {\bibfnamefont {A.~A.}\ \bibnamefont
  {Mostofi}}, \bibinfo {author} {\bibfnamefont {J.~R.}\ \bibnamefont {Yates}},
  \bibinfo {author} {\bibfnamefont {I.}~\bibnamefont {Souza}},\ and\ \bibinfo
  {author} {\bibfnamefont {D.}~\bibnamefont {Vanderbilt}},\ }\bibfield  {title}
  {\bibinfo {title} {Maximally localized {Wannier} functions: Theory and
  applications},\ }\href {https://doi.org/10.1103/RevModPhys.84.1419}
  {\bibfield  {journal} {\bibinfo  {journal} {Rev. Mod. Phys.}\ }\textbf
  {\bibinfo {volume} {84}},\ \bibinfo {pages} {1419} (\bibinfo {year}
  {2012})}\BibitemShut {NoStop}%
\bibitem [{\citenamefont {Hukushima}\ and\ \citenamefont
  {Nemoto}(1996)}]{PTMC}%
  \BibitemOpen
  \bibfield  {author} {\bibinfo {author} {\bibfnamefont {K.}~\bibnamefont
  {Hukushima}}\ and\ \bibinfo {author} {\bibfnamefont {K.}~\bibnamefont
  {Nemoto}},\ }\bibfield  {title} {\bibinfo {title} {Exchange {Monte} {Carlo}
  method and application to spin glass simulations},\ }\href
  {https://doi.org/10.1143/JPSJ.65.1604} {\bibfield  {journal} {\bibinfo
  {journal} {J. Phys. Soc. Jpn.}\ }\textbf {\bibinfo {volume} {65}},\ \bibinfo
  {pages} {1604} (\bibinfo {year} {1996})}\BibitemShut {NoStop}%
\bibitem [{\citenamefont {Metropolis}\ and\ \citenamefont
  {Ulam}(1949)}]{Metropolis1949}%
  \BibitemOpen
  \bibfield  {author} {\bibinfo {author} {\bibfnamefont {N.}~\bibnamefont
  {Metropolis}}\ and\ \bibinfo {author} {\bibfnamefont {S.}~\bibnamefont
  {Ulam}},\ }\bibfield  {title} {\bibinfo {title} {The {Monte} {Carlo}
  method},\ }\href
  {https://www.tandfonline.com/doi/abs/10.1080/01621459.1949.10483310}
  {\bibfield  {journal} {\bibinfo  {journal} {J. Am. Stat. Assoc.}\ }\textbf
  {\bibinfo {volume} {44}},\ \bibinfo {pages} {335} (\bibinfo {year}
  {1949})}\BibitemShut {NoStop}%
\bibitem [{\citenamefont {Mertens}\ \emph {et~al.}(2023)\citenamefont
  {Mertens}, \citenamefont {Mönkebüscher}, \citenamefont {Parlak},
  \citenamefont {Boix-Constant}, \citenamefont {Mañas-Valero}, \citenamefont
  {Matzer}, \citenamefont {Adhikari}, \citenamefont {Bonanni}, \citenamefont
  {Coronado}, \citenamefont {Kalashnikova}, \citenamefont {Bossini},\ and\
  \citenamefont {Cinchetti}}]{Mertens_2023}%
  \BibitemOpen
  \bibfield  {author} {\bibinfo {author} {\bibfnamefont {F.}~\bibnamefont
  {Mertens}}, \bibinfo {author} {\bibfnamefont {D.}~\bibnamefont
  {Mönkebüscher}}, \bibinfo {author} {\bibfnamefont {U.}~\bibnamefont
  {Parlak}}, \bibinfo {author} {\bibfnamefont {C.}~\bibnamefont
  {Boix-Constant}}, \bibinfo {author} {\bibfnamefont {S.}~\bibnamefont
  {Mañas-Valero}}, \bibinfo {author} {\bibfnamefont {M.}~\bibnamefont
  {Matzer}}, \bibinfo {author} {\bibfnamefont {R.}~\bibnamefont {Adhikari}},
  \bibinfo {author} {\bibfnamefont {A.}~\bibnamefont {Bonanni}}, \bibinfo
  {author} {\bibfnamefont {E.}~\bibnamefont {Coronado}}, \bibinfo {author}
  {\bibfnamefont {A.~M.}\ \bibnamefont {Kalashnikova}}, \bibinfo {author}
  {\bibfnamefont {D.}~\bibnamefont {Bossini}},\ and\ \bibinfo {author}
  {\bibfnamefont {M.}~\bibnamefont {Cinchetti}},\ }\bibfield  {title} {\bibinfo
  {title} {Ultrafast coherent thz lattice dynamics coupled to spins in the van
  der waals antiferromagnet {FePS}$_3$},\ }\href
  {https://onlinelibrary.wiley.com/doi/abs/10.1002/adma.202208355} {\bibfield
  {journal} {\bibinfo  {journal} {Adv. Mater.}\ }\textbf {\bibinfo {volume}
  {35}},\ \bibinfo {pages} {2208355} (\bibinfo {year} {2023})}\BibitemShut
  {NoStop}%
\bibitem [{\citenamefont {Lan\ifmmode~\mbox{\c{c}}\else \c{c}\fi{}on}\ \emph
  {et~al.}(2016)\citenamefont {Lan\ifmmode~\mbox{\c{c}}\else \c{c}\fi{}on},
  \citenamefont {Walker}, \citenamefont {Ressouche}, \citenamefont {Ouladdiaf},
  \citenamefont {Rule}, \citenamefont {McIntyre}, \citenamefont {Hicks},
  \citenamefont {R\o{}nnow},\ and\ \citenamefont {Wildes}}]{Lan2016}%
  \BibitemOpen
  \bibfield  {author} {\bibinfo {author} {\bibfnamefont {D.}~\bibnamefont
  {Lan\ifmmode~\mbox{\c{c}}\else \c{c}\fi{}on}}, \bibinfo {author}
  {\bibfnamefont {H.~C.}\ \bibnamefont {Walker}}, \bibinfo {author}
  {\bibfnamefont {E.}~\bibnamefont {Ressouche}}, \bibinfo {author}
  {\bibfnamefont {B.}~\bibnamefont {Ouladdiaf}}, \bibinfo {author}
  {\bibfnamefont {K.~C.}\ \bibnamefont {Rule}}, \bibinfo {author}
  {\bibfnamefont {G.~J.}\ \bibnamefont {McIntyre}}, \bibinfo {author}
  {\bibfnamefont {T.~J.}\ \bibnamefont {Hicks}}, \bibinfo {author}
  {\bibfnamefont {H.~M.}\ \bibnamefont {R\o{}nnow}},\ and\ \bibinfo {author}
  {\bibfnamefont {A.~R.}\ \bibnamefont {Wildes}},\ }\bibfield  {title}
  {\bibinfo {title} {Magnetic structure and magnon dynamics of the
  quasi-two-dimensional antiferromagnet {FePS}$_{3}$},\ }\href
  {https://link.aps.org/doi/10.1103/PhysRevB.94.214407} {\bibfield  {journal}
  {\bibinfo  {journal} {Phys. Rev. B}\ }\textbf {\bibinfo {volume} {94}},\
  \bibinfo {pages} {214407} (\bibinfo {year} {2016})}\BibitemShut {NoStop}%
\bibitem [{\citenamefont {Cenker}\ \emph {et~al.}(2022)\citenamefont {Cenker},
  \citenamefont {Sivakumar}, \citenamefont {Xie}, \citenamefont {Miller},
  \citenamefont {Thijssen}, \citenamefont {Liu}, \citenamefont {Dismukes},
  \citenamefont {Fonseca}, \citenamefont {Anderson}, \citenamefont {Zhu},
  \citenamefont {Roy}, \citenamefont {Xiao}, \citenamefont {Chu}, \citenamefont
  {Cao},\ and\ \citenamefont {Xu}}]{Cenker2022}%
  \BibitemOpen
  \bibfield  {author} {\bibinfo {author} {\bibfnamefont {J.}~\bibnamefont
  {Cenker}}, \bibinfo {author} {\bibfnamefont {S.}~\bibnamefont {Sivakumar}},
  \bibinfo {author} {\bibfnamefont {K.}~\bibnamefont {Xie}}, \bibinfo {author}
  {\bibfnamefont {A.}~\bibnamefont {Miller}}, \bibinfo {author} {\bibfnamefont
  {P.}~\bibnamefont {Thijssen}}, \bibinfo {author} {\bibfnamefont
  {Z.}~\bibnamefont {Liu}}, \bibinfo {author} {\bibfnamefont {A.}~\bibnamefont
  {Dismukes}}, \bibinfo {author} {\bibfnamefont {J.}~\bibnamefont {Fonseca}},
  \bibinfo {author} {\bibfnamefont {E.}~\bibnamefont {Anderson}}, \bibinfo
  {author} {\bibfnamefont {X.}~\bibnamefont {Zhu}}, \bibinfo {author}
  {\bibfnamefont {X.}~\bibnamefont {Roy}}, \bibinfo {author} {\bibfnamefont
  {D.}~\bibnamefont {Xiao}}, \bibinfo {author} {\bibfnamefont {J.-H.}\
  \bibnamefont {Chu}}, \bibinfo {author} {\bibfnamefont {T.}~\bibnamefont
  {Cao}},\ and\ \bibinfo {author} {\bibfnamefont {X.}~\bibnamefont {Xu}},\
  }\bibfield  {title} {\bibinfo {title} {Reversible strain-induced magnetic
  phase transition in a van der {Waals} magnet},\ }\href
  {https://www.nature.com/articles/s41565-021-01052-6} {\bibfield  {journal}
  {\bibinfo  {journal} {Nat. Nanotechnol.}\ }\textbf {\bibinfo {volume} {17}},\
  \bibinfo {pages} {256} (\bibinfo {year} {2022})}\BibitemShut {NoStop}%
\bibitem [{\citenamefont {Wang}\ \emph {et~al.}(2020)\citenamefont {Wang},
  \citenamefont {Wang}, \citenamefont {Liang}, \citenamefont {Ma},
  \citenamefont {Xu}, \citenamefont {Liu}, \citenamefont {Zhang}, \citenamefont
  {Admasu}, \citenamefont {Cheong}, \citenamefont {Wang}, \citenamefont {Chen},
  \citenamefont {Liu}, \citenamefont {Cheng}, \citenamefont {Ji},\ and\
  \citenamefont {Miao}}]{Wang2020}%
  \BibitemOpen
  \bibfield  {author} {\bibinfo {author} {\bibfnamefont {Y.}~\bibnamefont
  {Wang}}, \bibinfo {author} {\bibfnamefont {C.}~\bibnamefont {Wang}}, \bibinfo
  {author} {\bibfnamefont {S.-J.}\ \bibnamefont {Liang}}, \bibinfo {author}
  {\bibfnamefont {Z.}~\bibnamefont {Ma}}, \bibinfo {author} {\bibfnamefont
  {K.}~\bibnamefont {Xu}}, \bibinfo {author} {\bibfnamefont {X.}~\bibnamefont
  {Liu}}, \bibinfo {author} {\bibfnamefont {L.}~\bibnamefont {Zhang}}, \bibinfo
  {author} {\bibfnamefont {A.~S.}\ \bibnamefont {Admasu}}, \bibinfo {author}
  {\bibfnamefont {S.-W.}\ \bibnamefont {Cheong}}, \bibinfo {author}
  {\bibfnamefont {L.}~\bibnamefont {Wang}}, \bibinfo {author} {\bibfnamefont
  {M.}~\bibnamefont {Chen}}, \bibinfo {author} {\bibfnamefont {Z.}~\bibnamefont
  {Liu}}, \bibinfo {author} {\bibfnamefont {B.}~\bibnamefont {Cheng}}, \bibinfo
  {author} {\bibfnamefont {W.}~\bibnamefont {Ji}},\ and\ \bibinfo {author}
  {\bibfnamefont {F.}~\bibnamefont {Miao}},\ }\bibfield  {title} {\bibinfo
  {title} {Strain-sensitive magnetization reversal of a van der {Waals}
  magnet},\ }\href {https://onlinelibrary.wiley.com/doi/10.1002/adma.202004533}
  {\bibfield  {journal} {\bibinfo  {journal} {Adv. Mater.}\ }\textbf {\bibinfo
  {volume} {32}},\ \bibinfo {pages} {2004533} (\bibinfo {year}
  {2020})}\BibitemShut {NoStop}%
\end{thebibliography}%

\newpage
\begin{appendix}
	\setcounter{figure}{0}
	\setcounter{table}{0}
	\renewcommand{\thefigure}{S\arabic{figure}}
	\renewcommand{\thetable}{S\arabic{table}}
	\renewcommand{\theequation}{S\arabic{equation}}
	\renewcommand{\tablename}{Table}
	\renewcommand{\figurename}{Fig.}

\titleformat*{\section}{\normalfont\Large\bfseries}
\section*{Supplemental Material for "Understanding the Ising zigzag antiferromagnetism of FePS$_3$ and FePSe$_3$ monolayers"}

\renewcommand\arraystretch{1.3}
\begin{table}[H]
	\centering
	\caption{Relative total energies (meV/f.u.) of the 3$d^{5\uparrow}L_{z+}^{1\downarrow}$ ground state against the 3$d^{5\uparrow}a_{1g}^{1\downarrow}$ state using different cutoff energies (eV), $k$-meshes, and $U$ values (eV) for FePS$_3$ monolayer by GGA + SOC + $U$ calculations.}
	\begin{tabular}{c@{\hskip3mm}c@{\hskip3mm}c@{\hskip3mm}c@{\hskip3mm}c@{\hskip3mm}}
		\hline\hline
		$E_{\rm cut}$ (eV) & $k$-mesh    & $U$ (eV)  & 3$d^{5\uparrow}L_{z+}^{1\downarrow}$  & 3$d^{5\uparrow}a_{1g}^{1\downarrow}$   \\ 
		\hline
		450 & 6$\times$6$\times$1 & 3.6 & 0  & 125.1  \\
		450 & 9$\times$9$\times$1 & 3.6 & 0   & 125.1 \\
		550 & 6$\times$6$\times$1 & 3.6 & 0  & 124.9 \\ \hline
		450 & 6$\times$6$\times$1 & 4.0 & 0   & 113.8\\  
		\hline\hline
	\end{tabular}
	\label{U}
\end{table}
\renewcommand\arraystretch{1.3}
\begin{table}[H]
	\centering
	\caption{Relative total energies $\Delta$\textit{E} (meV/f.u.), local spin and orbital moments ($\mu_{\rm B}$) and the derived three exchange parameters (meV) for FePS$_3$ and FePSe$_3$ monolayers both in the 3$d^{5\uparrow}$$L_{z+}^{1\downarrow}$ ground state by GGA + SOC + $U$ calculations with $U$ = 4.0 eV.}
	\begin{tabular}{c@{\hskip5mm}c@{\hskip5mm}c@{\hskip5mm}r@{\hskip5mm}r@{\hskip5mm}}
		\hline\hline
		Systems  & States  & $\Delta$\textit{E} & Fe$_{\rm spin}$  & Fe$_{\rm orb}$    \\ \hline
		\multirow{5}{*}{FePS$_3$} &	zigzag AF  &  0       &  $\pm$3.54  &  $\pm$0.77      \\
		&	FM      &  15.6           &  3.57       &  0.74          \\
		&	N$\acute{e}$el AF  &  24.4         &  $\pm$3.53  &  $\pm$0.85      \\
		&	stripe AF  &  32.1        &  $\pm$3.55  &  $\pm$0.79     \\
		&	\multicolumn{1}{c}{\textit{J}$_{1}$=2.56}    & \multicolumn{2}{c}{\textit{J}$_{2}$=--0.25} &    \multicolumn{1}{c}{\textit{J}$_{3}$=--1.83}    \\  \hline
		\multirow{5}{*}{FePSe$_3$} & zigzag AF  &  0       &  $\pm$3.51  &  $\pm$0.75      \\
		&N$\acute{e}$el AF  &  17.4         &  $\pm$3.49  &  $\pm$0.83      \\
		&FM      &  19.1           &  3.54       &  0.72          \\
		&stripe AF  &  29.4        &  $\pm$3.52  &  $\pm$0.76     \\
		&\multicolumn{1}{c}{\textit{J}$_{1}$=1.73}    & \multicolumn{2}{c}{\textit{J}$_{2}$=--0.22} &    \multicolumn{1}{c}{\textit{J}$_{3}$=--1.87}    \\
		\hline\hline
	\end{tabular}
	\label{J}
\end{table}
\begin{figure}[H]
	\includegraphics[width=7cm]{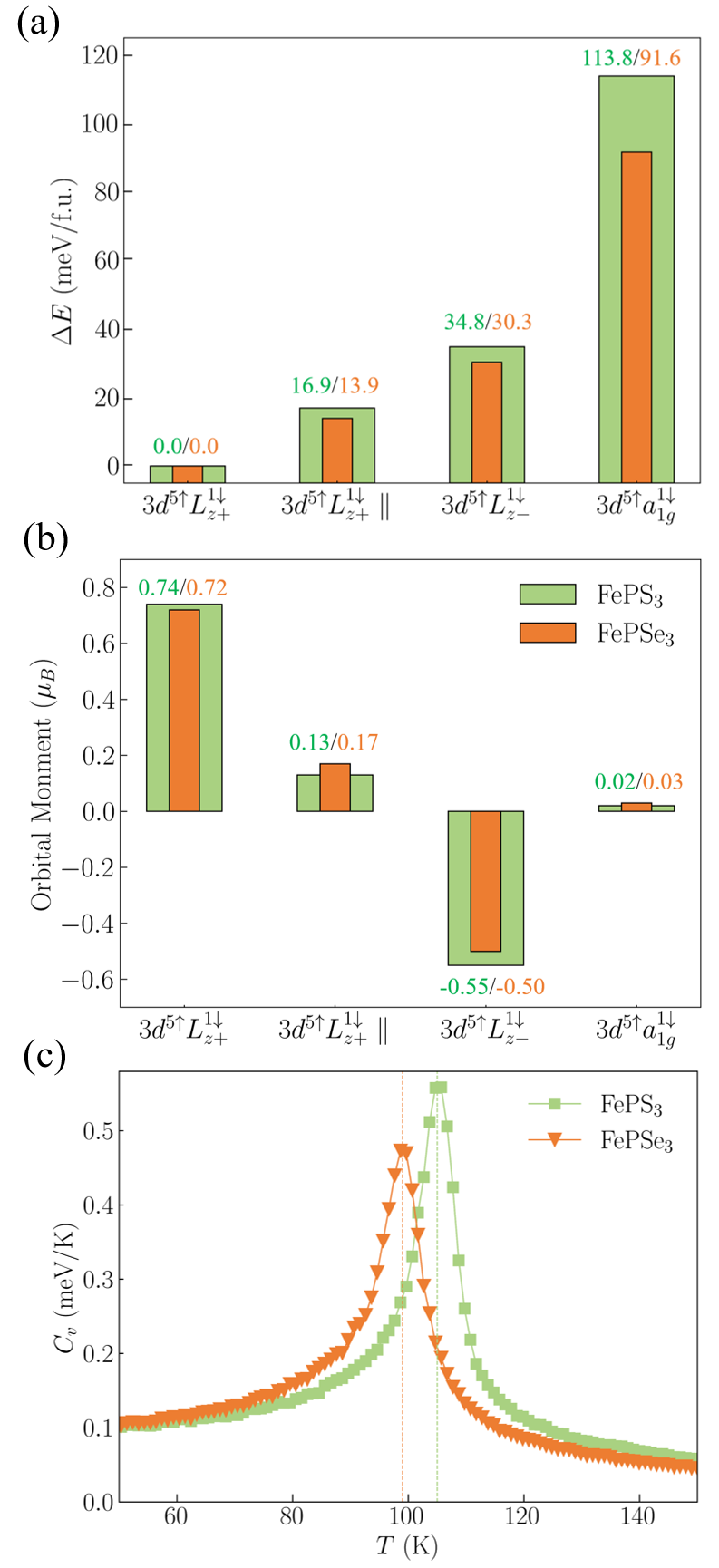}
	\centering
	\caption{(a) Relative total energies $\Delta$\textit{E} (meV/f.u.), and (b) orbital moments ($\mu_{\rm B}$) of FePS$_3$ and FePSe$_3$ monolayers in different spin-orbital states by GGA + SOC + $U$ calculations with $U$ = 4.0 eV. (c) PTMC simulations of the magnetic specific heat of FePS$_3$ and FePSe$_3$ monolayers.}
\end{figure}

\begin{figure}[H]
	\includegraphics[width=7cm]{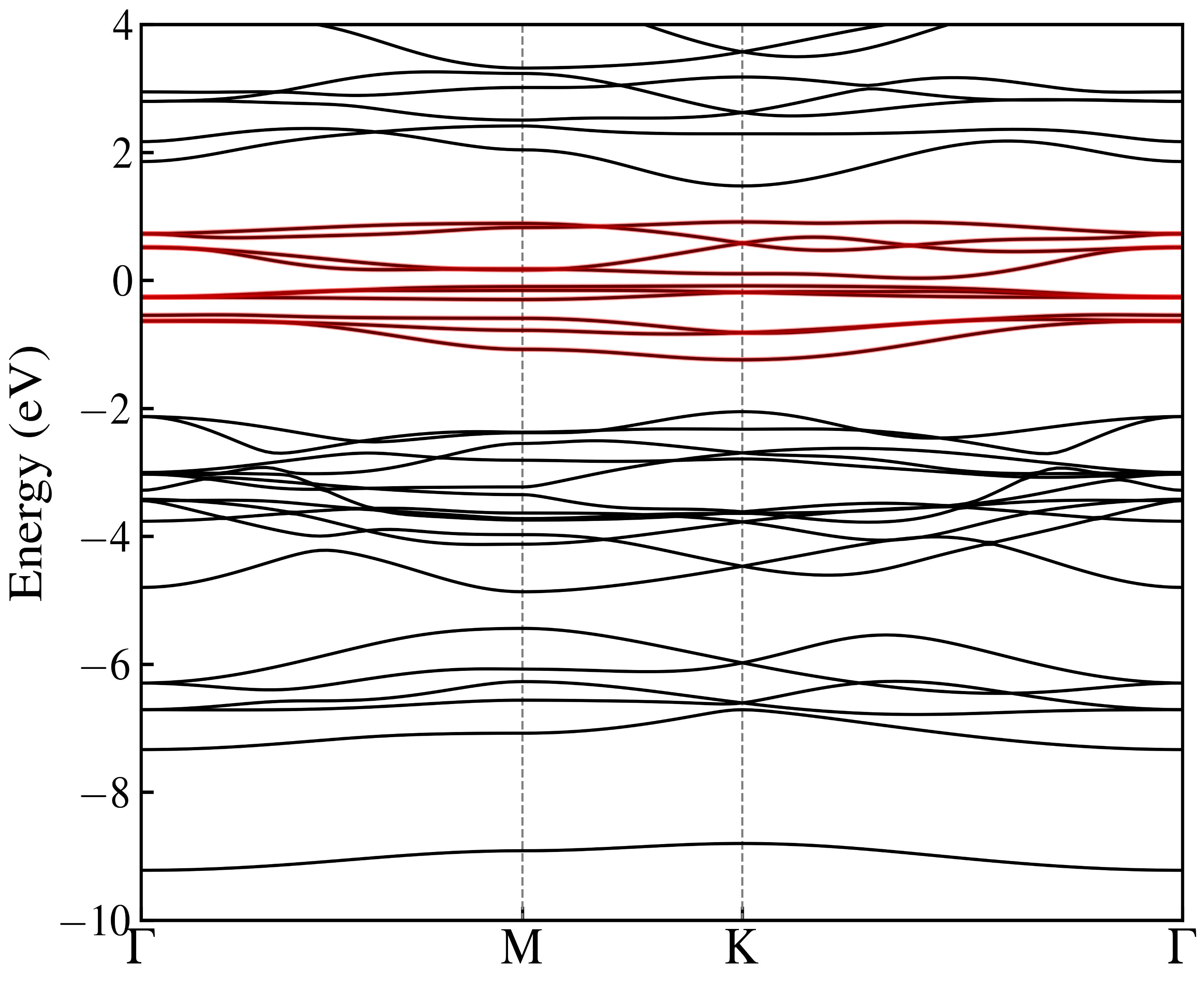}
	\centering
	\caption{The calculated DFT (black solid lines) and Wannier-interpolated (red dashed lines) band structures of FePS$_3$ monolayer.}
	\label{strains}
\end{figure}

\begin{figure}[H]
	\includegraphics[width=8cm]{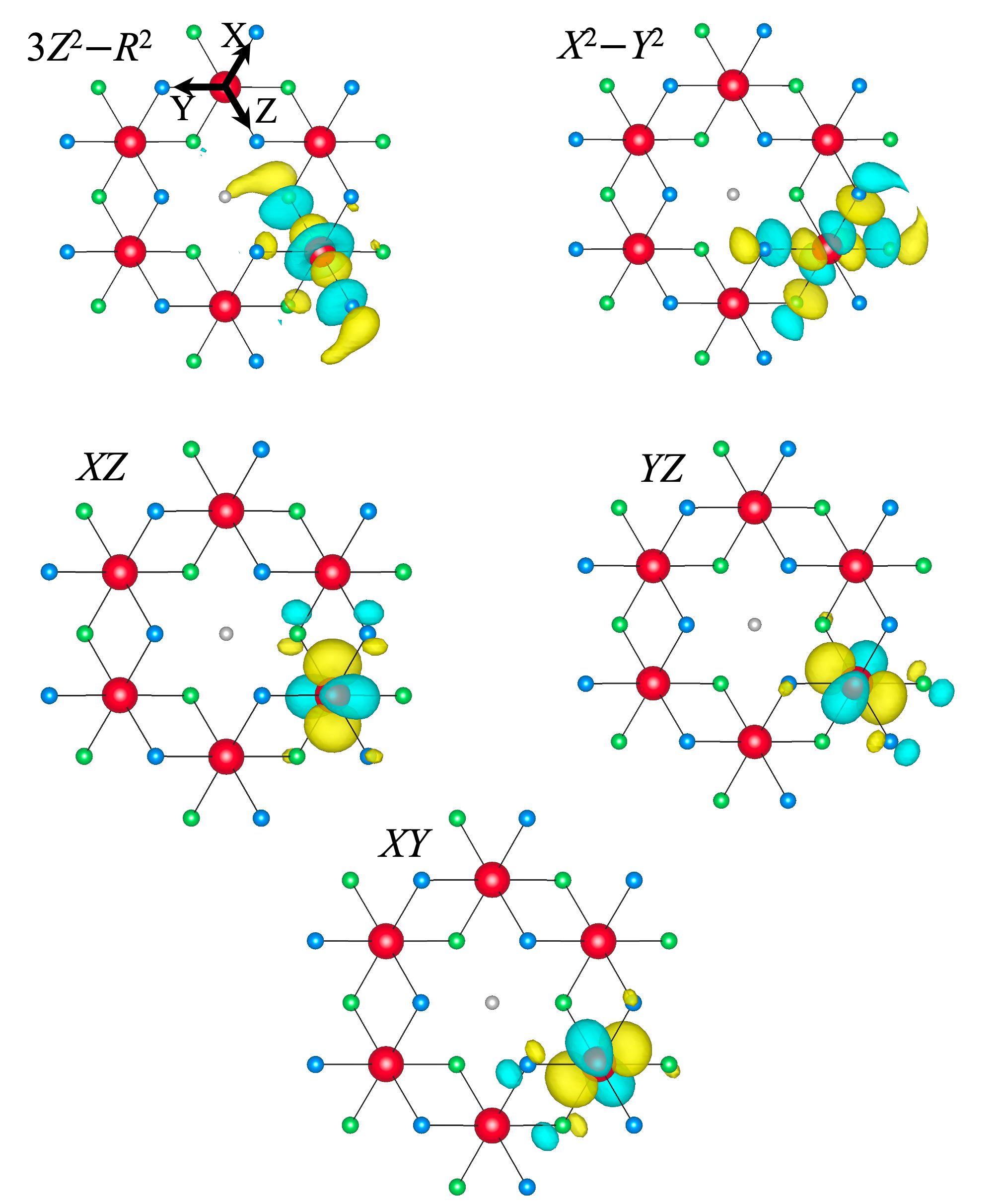}
	\centering
	\caption{Contour-surface plots of the Fe 3$d$ Wannier functions. For all plots, we choose an isosurface level of $\pm$2.3 (yellow for positive values and cyan for negative values).}
\end{figure}
\begin{figure}[H]
	\includegraphics[width=8cm]{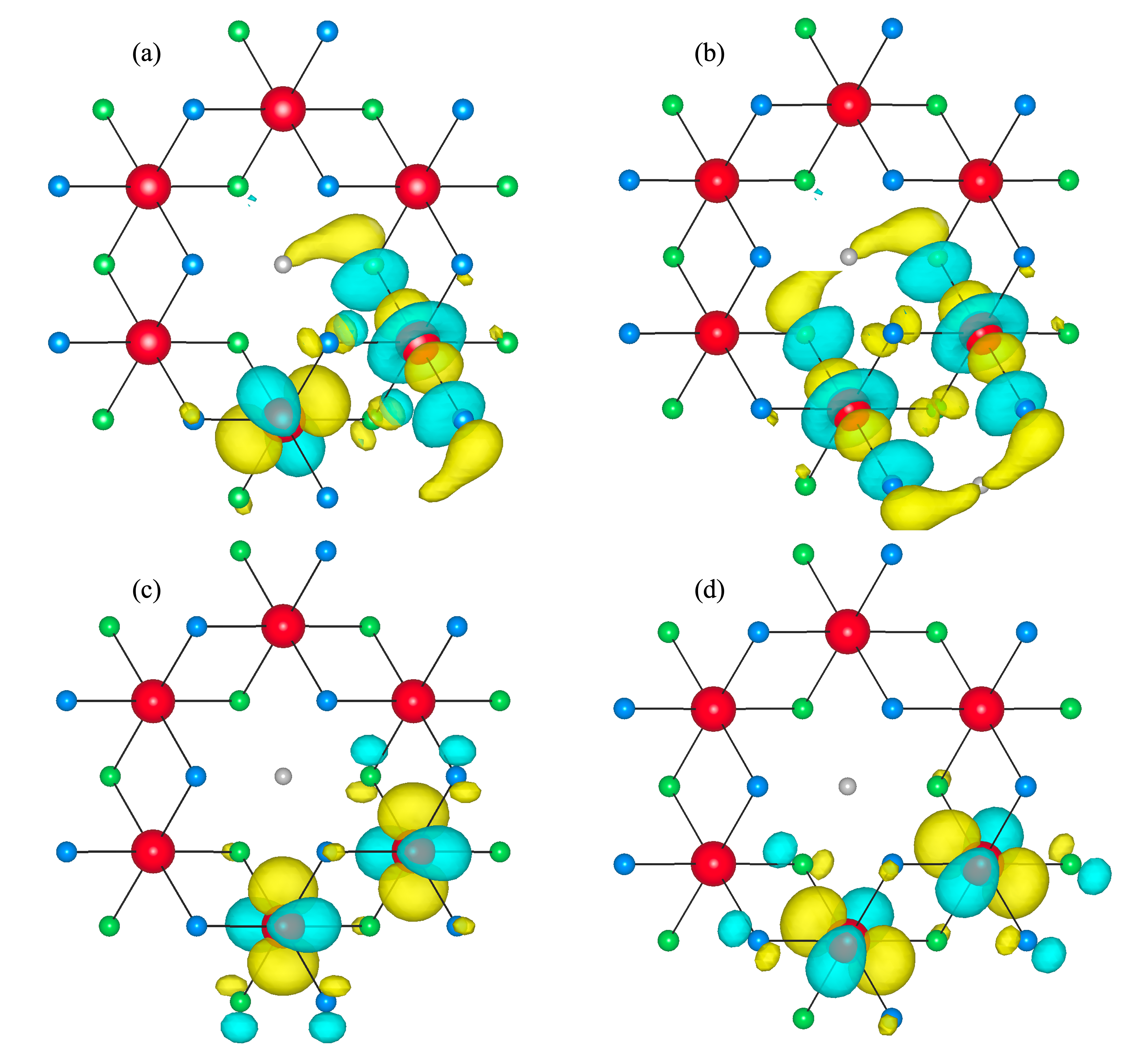}
	\centering
	\caption{Schematic plots of 1NN hopping channels in FePS$_3$ monolayer.}
\end{figure}
\begin{figure}[H]
	\includegraphics[width=5cm]{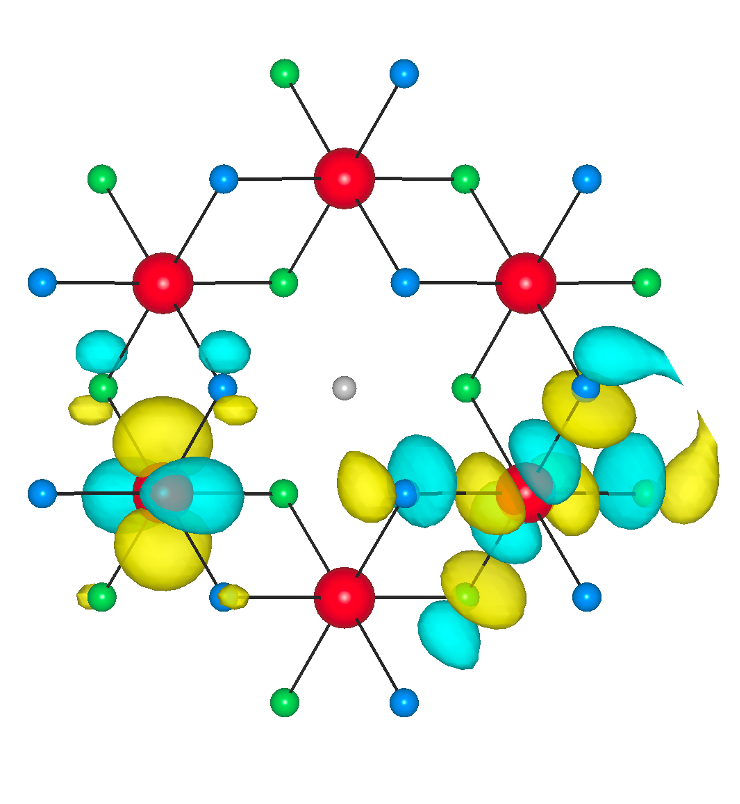}
	\centering
	\caption{Schematic plot of 2NN hopping channel in FePS$_3$ monolayer.}
\end{figure}
\begin{figure}[H]
	\includegraphics[width=3.8cm]{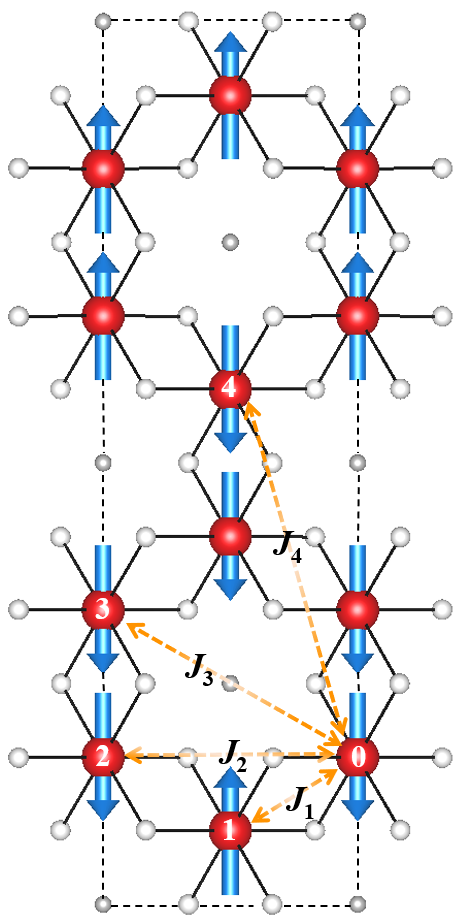}
	\centering
	\caption{The double-stripe AF structure of FePS$_3$ monolayer marked with four exchange parameters.}
	\label{J4}
\end{figure}
\begin{figure}[H]
	\includegraphics[width=9cm]{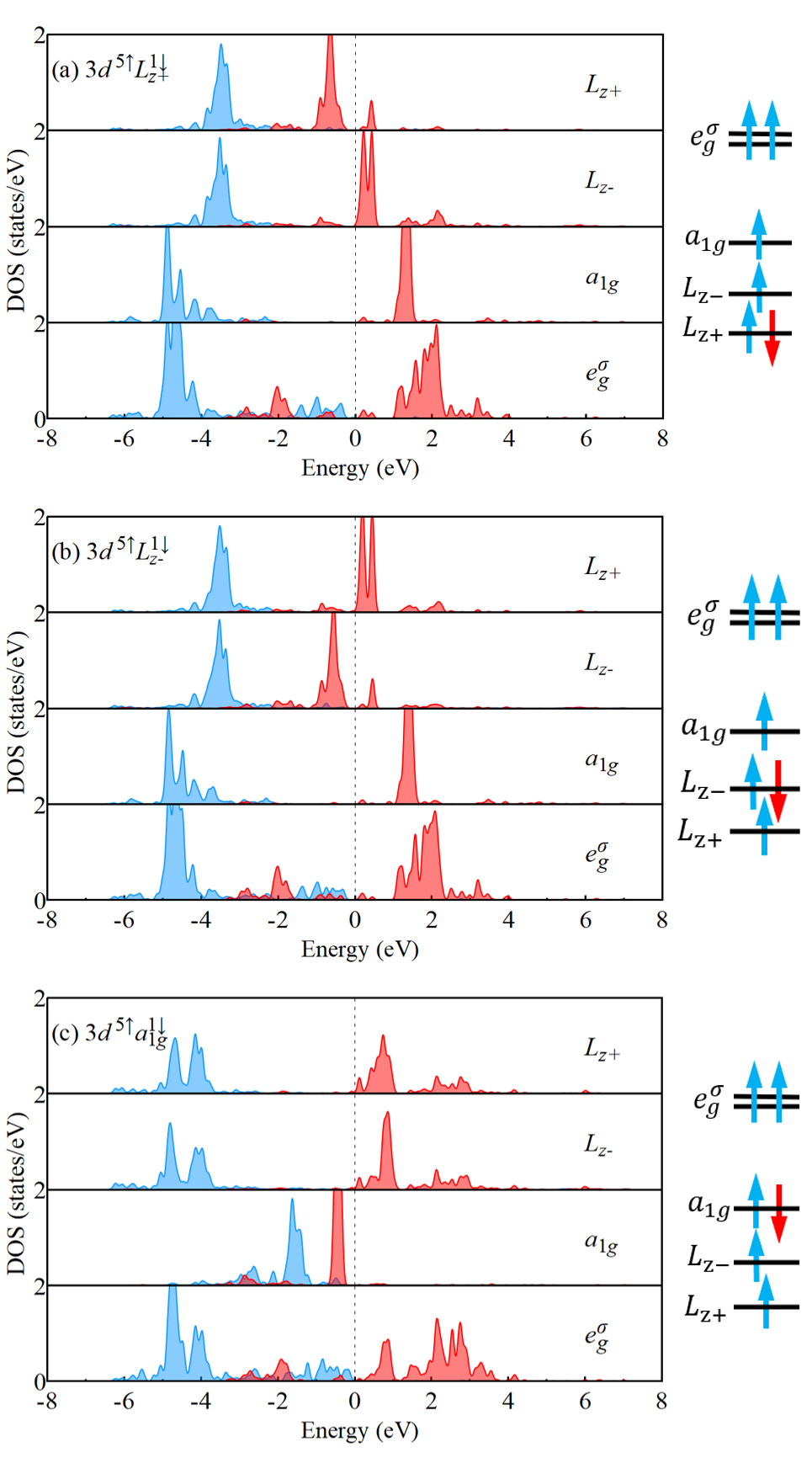}
	\centering
	\caption{The DOS results of FePSe$_3$ monolayer in (a) the 3$d^{5\uparrow}$$L_{z+}^{1\downarrow}$ ground state, (b) 3$d^{5\uparrow}$$L_{z-}^{1\downarrow}$ and (c) 3$d^{5\uparrow}$$a_{1g}^{1\downarrow}$ states by the GGA + SOC + $U$ calculations with $U$ = 2.7 eV.  The blue (red) curves stand for the up (down) spin channel. The Fermi level is set at zero energy.}
	\label{FePSe3_GGASUSOC}
\end{figure}
%


\end{appendix}

\end{document}